\documentclass[superscriptaddress,nofootinbib,twocolumn,prb]{revtex4-1}
\usepackage{amsmath,parskip,graphicx,acro,array}
\usepackage{siunitx}  

\newcommand{\e}{\varepsilon}
\newcommand{\F}{\mathcal{F}}

\renewcommand{\(}{\left(}
\renewcommand{\)}{\right)}
\renewcommand{\vec}[1]{\mathbf{#1}}

\newcommand{\pdif}[2]{\frac{{\rm \partial}#1}{{{\rm \partial}#2}}}

\begin{document}

\title{Solvent fluctuations around solvophobic, solvophilic and patchy nanostructures and the accompanying solvent mediated interactions}
\author{Blesson Chacko}
\affiliation{Department of Mathematical Sciences, Loughborough University, Loughborough, LE11 3TU, UK}
\author{Robert Evans}
\affiliation{H. H. Wills Physics Laboratory, University of Bristol, Bristol, BS8 1TL, UK}
\author{Andrew J. Archer}
\affiliation{Department of Mathematical Sciences, Loughborough University, Loughborough, LE11 3TU, UK}
\date{\today}

\begin{abstract}
Using classical density functional theory (DFT) we calculate the density profile $\rho(\vec{r})$ and local compressibility $\chi(\vec{r})$ of a simple liquid solvent in which a pair of blocks with (microscopic) rectangular cross-section are immersed. We consider blocks that are solvophobic, solvophilic and also ones that have both solvophobic and solvophilic patches. Large values of $\chi(\vec{r})$ correspond to regions in space where the liquid density is fluctuating most strongly. We seek to elucidate how enhanced density fluctuations correlate with the solvent mediated force between the blocks, as the distance between the blocks and the chemical potential of the liquid reservoir vary. For sufficiently solvophobic blocks, at small block separations and small deviations from bulk gas-liquid coexistence, we observe a strongly attractive (near constant) force, stemming from capillary evaporation to form a low density gas-like intrusion between the blocks. The accompanying $\chi(\vec{r})$ exhibits structure which reflects the incipient gas-liquid interfaces that develop. We argue that our model system provides a means to understanding the basic physics of solvent mediated interactions between nanostructures, and between objects such as proteins in water, that possess hydrophobic and hydrophilic patches.
\end{abstract}

\maketitle

\section{Introduction}

Understanding the properties of water near hydrophobic surfaces continues to attract attention across several different disciplines,\cite{chandler2005interfaces, berne2009dewetting} ranging from the design  of self-cleaning materials \cite{ueda2013emerging, aytug2015monolithic} to biological self-assembly and protein interactions.\cite{ball2008water} Likewise, understanding the (water mediated) interactions between hydrophobic and hydrophilic entities is important in many areas of physical chemistry and chemical physics. In a recent article, Kandu\v{c} {\it et al.}\cite{kanduc2016water} survey the field and describe informatively how the behaviour of soft-matter at the nano-scale depends crucially on surface properties and outline the key role played by water mediated interactions in many technological and biological processes. These include colloid science, where altering the surface chemistry can change enormously the effective interactions, e.g. those preventing aggregation, and biological matter where effective membrane-membrane interactions can be important in biological processes. 

In attempting to ascertain the nature of effective interactions, it is crucial to know whether a certain substrate, or entity, is hydrophilic or hydrophobic. For a macroscopic (planar) substrate the degree of hydrophobicity is measured by Young's contact angle $\theta$. A strongly hydrophobic surface, such as a self-assembled monolayer (SAM), paraffin or hydrocarbon, can have $\theta > 120^\circ$, while a strongly hydrophilic surface can often correspond to complete wetting, i.e.\ $\theta=0$, meaning a water drop spreads across the whole surface. However, in the majority of systems encountered in the physical chemistry of colloids, in nanoscience and in situations pertinent to biological systems, the entities immersed in water do not have a macroscopic surface area. Thus it is important to ask to what extent ideas borrowed from a macroscopic (capillarity) description, which simply balance bulk (volume) and surface (area) contributions to the total grand potential but which make specific predictions for the effective interaction between two immersed macroscopic hydrophobic entities, remain valid at the nanoscale.  For example, Huang {\it et al.}~\cite{huang2003dewetting} consider the phenomenon of capillary evaporation of SPC water between two hydrophobic oblate (ellipsoidal) plates. These authors discuss the validity of the macroscopic formula at which evaporation occurs and the form of the solvent mediated force between the plates. More recently, Jabes {\it et al.}~\cite{jabes2016universal} investigate the solvent-induced interactions for SPC/E water between curved hydrophobes; they consider the influence of different types of confining geometry and conclude that macroscopic thermodynamic (capillarity) arguments work surprisingly well, even at length scales corresponding to a few molecular (water) diameters. The survey article Ref.~\onlinecite{kanduc2016water} emphasises the usefulness of capillarity ideas for analysing water mediated forces between two entities that have different adsorbing strengths.

Such observations raise the general physics question as to how well should one expect capillarity arguments to work for nanoscale entities immersed in an arbitrary solvent. Are these observations specific to water? This seems most unlikely. In this paper we argue that insight into fundamental aspects of solvent mediated interactions, particularly those pertaining to solvophobes, are best addressed by considering the effective, solvent mediated interactions between nanostructures immersed in a simple Lennard-Jones (LJ) liquid. By focusing on a model liquid with much simpler intermolecular forces than those in water, one can investigate more easily and more systematically the underlying physics, e.g.\ the length scales relevant for phenomena such as capillary evaporation and how these determine the effective interactions.
    
A second, closely related, aspect of our present study is concerned with the strength and range of density fluctuations in water close to hydrophobic substrates. It is now accepted that for water near a macroscopic strongly hydrophobic substrate the local number density of the water is reduced below that in bulk for the first one or two adsorbed molecular layers. Accompanying this reduction in local density there is growing evidence for a substantial increase in fluctuations in the local number density; these increase for increasing water contact angle. An illuminating review~\cite{jamadagni2011hydrophobicity} surveys the field up to 2011, describing earlier work on density fluctuations, from the groups of Garde, Hummer, and Chandler. The basic idea of Garde and co-workers is that a large value of some, appropriately defined, local compressibility reflects the strength of density fluctuations in the neighbourhood of the substrate and should provide a quantitative measure of the degree of hydrophobicity of the hydrophobic entity.\cite{jamadagni2011hydrophobicity} The idea is appealing. However, even for a macroscopic planar substrate, there are problems in deciding upon the appropriate measure. Once again, this issue is not specific to water. If strong fluctuations occur at hydrophobic surfaces one should also expect these to occur at solvophobic surfaces, for similar values of chemical potential deviation from bulk coexistence. In other words, pronounced fluctuations cannot be specific to water near hydrophobic substrates. This argument was outlined recently.\cite{evans2015local,evans2015quantifying}

Evans and Stewart~\cite{evans2015local} discuss the merits of various different quantities that measure surface fluctuations. They argue that the compressibility $\chi(\vec{r})$, defined as the derivative of the equilibrium density, $\rho(\vec{r})$, with respect to the chemical potential $\mu$ at fixed temperature $T$:
\begin{equation}\label{eq:chi}
	\chi(\vec{r}) \equiv \( \pdif{ \rho(\vec{r}) }{ \mu } \)_T
\end{equation}
provides the most natural and useful measure for quantifying the local fluctuations in an inhomogeneous liquid. This quantity was introduced much earlier,\cite{tarazona1982long,evans1989wetting,stewart2012phase} in studies of wetting/drying and confined fluids and is, of course, calculated in the grand canonical ensemble. The usual isothermal compressibility~\cite{hansen2013theory} $\kappa_T=\chi_b/\rho_b^2$, where $\chi_b\equiv(\partial\rho_b/\partial\mu)_T$ is the bulk value of the compressibility; recall that $\chi_b\to\infty$ on approaching the bulk fluid critical point. Note that $\chi(\vec{r})$ can be expressed\cite{evans2015quantifying} as the correlator
\begin{equation}\label{eq:chi_2}
	\chi(\vec{r}) = \beta \langle N \hat{\rho}(\vec{r}) -\langle N\rangle \langle \hat{\rho}(\vec{r})\rangle\rangle
\end{equation}
where $\beta=(k_BT)^{-1}$, $\hat{\rho}(\vec{r})$ is the particle density operator, $N= \int \hat{\rho}(\vec{r}) \mathrm{d}\vec{r}$ is the number of particles and $\langle \cdots \rangle$ denotes a grand canonical average. Thus $\langle \hat{\rho}(\vec{r})\rangle =\rho(\vec{r})$ and $\langle N \rangle$ is the average number of particles. Clearly $\chi(\vec{r})$ correlates the local number density at $\vec{r}$ with the total number of particles in the system. The measures of $\chi(\vec{r})$ introduced by other authors\cite{acharya2010mapping,willard2014molecular} are designed for molecular dynamics computations which are performed in the canonical ensemble rather than in the grand canonical ensemble. The latter is more appropriate for adsorption studies.

Using DFT, Evans and Stewart~\cite{evans2015local} calculated $\chi(z)$ defined by Eq.\ \eqref{eq:chi} for LJ liquids near planar substrates, with the wall at $z=0$. They investigated substrates which ranged from neutral $(\theta\approx90^\circ)$ to very solvophobic $(\theta\approx170^\circ)$ and found that this quantity is enhanced over bulk, exhibiting a peak for $z$ within one or two atomic diameters of the substrate. The height of the peak increased significantly as $\theta$ increased and the substrate became more solvophobic. In subsequent investigations, using Grand Canonical Monte Carlo (GCMC) for SPC/E water~\cite{evans2015quantifying} and GCMC plus DFT for a LJ liquid~\cite{evans2016critical} at model solvophobic substrates, it was observed the the maximum in $\chi(z)$ increases rapidly as the strength of the wall-fluid attraction is reduced, thereby increasing $\theta$ towards $180^\circ$, i.e.\ towards complete drying. For different choices of wall-fluid potentials the drying transition is continuous (critical) and the thickness of the intruding gas-like layer as well as the maximum in $\chi(z)$ diverge as $\cos\theta \to -1$.~\cite{evans2015quantifying,evans2016critical} These observations pertain to the liquid at coexistence, where $\mu = \mu_\mathrm{coex}^+$.

Much is made in the literature concerning the depleted local density and accompanying  enhanced surface fluctuations of water at hydrophobic surfaces as arising from the particular properties of water, namely the hydrogen-bonding and the open tetrahedrally coordinated liquid structure, which is said to be disrupted by the presence of large enough hydrophobic objects. However, following from Refs.~\onlinecite{evans2015local,evans2015quantifying} we show here that much of this phenomenology is also observed when a simple LJ like solvent that is near to bulk gas-liquid phase coexistence is in contact with solvophobic objects. The particular entities we consider are i) planar surfaces of infinite area and ii) long blocks with a finite rectangular-cross-section. For these objects to be solvophobic, we treat them as being composed of particles to which the solvent particles are attracted weakly, compared to the strength of the attraction between solvent particles themselves. The contact angle of the solvent liquid at the planar solvophobic substrate considered here is $\theta\approx144^\circ$. We also consider the behaviour at solvophilic objects, for which the contact angle at the corresponding planar substrate is $\theta\approx44^\circ$.

The simple LJ like solvent we consider consists of particles with a hard-sphere pair interaction plus an additional attractive tail potential that decays $\sim r^{-6}$, where $r$ is the distance between the solvent particles. We use classical density functional theory (DFT),\cite{evans1979nature,evans1992density,hansen2013theory} treating the hard core interactions using the White-Bear version of fundamental measure theory (FMT),\cite{roth2002fundamental, roth2010fundamental} together with a mean-field treatment of the attractive interactions, to calculate the solvent density profile $\rho(\vec{r})$ and local compressibility $\chi(\vec{r})$. An advantage of using DFT is that having calculated $\rho(\vec{r})$, one then has access to all thermodynamic quantities including the various interfacial tensions. Calculating the grand potential as a function of the distance between the blocks yields the effective solvent mediated potential; minus the derivative of this quantity is the solvent mediated force between the blocks.

When both blocks are solvophobic {\em and} the liquid is at a state point near to bulk gas-liquid phase coexistence, we find that the solvent mediated force between these is strongly attractive at short distances due to the formation of a gas-like intrusion. Proximity to coexistence can be be quantified by the difference $\Delta\mu=\mu-\mu_\mathrm{coex}$, where $\mu_\mathrm{coex}$ is the value at bulk gas-liquid coexistence. For a slit pore consisting of two parallel surfaces of infinite extent that are sufficiently solvophobic, $\theta>90^\circ$, a first order transition, namely capillary evaporation, occurs as $\Delta\mu\to0$, corresponding to the stabilisation of the incipient gas phase in the slit of finite width.\cite{tarazona1987phase,evans1990fluids, gelb1999phase} The formation of the gas-like intrusion between the blocks that we consider here occurs at smaller $\Delta\mu$. This is not a genuine first order surface phase transition, owing to the finite size of the blocks. However, this phenomenon is intimately related to the capillary evaporation that occurs between parallel planar surfaces with both dimensions infinite. It turns into the genuine capillary evaporation phase transition as the height of our blocks is increased to $\infty$. Note that some authors in the water community, e.g.\ Refs.\ \onlinecite{berne2009dewetting}, \onlinecite{huang2003dewetting} and Remsing {\it et al.},\cite{remsing2015pathways} refer to this phenomenon as ``dewetting'', but given that this term is also used to describe a film of liquid on a single planar surface breaking up to form droplets, a network pattern or other structures,\cite{reiter1992dewetting, seemann2001dewetting, thiele2003open, thiele2010thin, archer2010dynamical} we prefer to use the more accurate term, capillary evaporation. The important matter of nomenclature was emphasised in a Faraday Discussion on hydrophobic and structured surfaces; see Refs.\ \onlinecite{luzar2010wetting,evans2010wetting}.

We also present results for the local compressibility $\chi(\vec{r})$ in the vicinity of the blocks. Maxima in $\chi(\vec{r})$ correspond to points in space where the density fluctuations are the greatest. We find that the formation of the gas-like intrusion between the hydrophobic blocks is associated with a local value of $\chi(\vec{r})$ that is much greater than the bulk value. However, we find that the solvent density fluctuations are not necessarily at points in space that one might initially expect. For example, when there is a gas-like intrusion, the value of $\chi(\vec{r})$ is larger at the entrance to the gap between the blocks, rather than in the centre of the gap.

We are not the first to use DFT to study liquids near corners and between surfaces. Bryk {\it et al.}~\cite{bryk2003depletion} calculated the solvent mediated (depletion) potential between a hard-sphere colloidal particle, immersed in a solvent of smaller hard-spheres, and planar substrates or geometrically structured substrates, including a right-angled wedge. They found that in the wedge geometry there is a strong attraction of the colloid to inner corners, but there is a free energy barrier repelling the colloid from an outer corner (edge) of a wedge. Hopkins {\it et al.} \cite{hopkins2009solvent} studied the solvent mediated interaction between a spherical (soft-core) particle, several times larger than the (soft-core) solvent particles, and a planar interface. They showed that when the binary solvent surrounding the large particle (colloid) is near to liquid-liquid phase coexistence, thick (wetting) films rich in the minority solvent species can form around it and on the interface. This has a profound effect on the solvent mediated potential, making it strongly attractive. A similar effect, due to proximity to liquid-liquid phase separation, was found for the solvent mediated potential between pairs of spherical colloidal particles.~\cite{hopkins2009solvent, archer2003solvent, archer2005solvent} Analogous effects arising from proximity to gas-liquid phase coexistence, i.e.\ when $\Delta\mu$ is small were found in a very recent study.~\cite{malijevsky2015bridging} Such investigations, studying the influence of proximity to bulk phase coexistence on the solvent mediated potential between pairs of spherical particles, provide insight regarding what one might expect in the cases studied here, namely pairs of hydrophobic, hydrophilic and patchy blocks.

The strong attractive forces between solvophobic objects, decreased local density and enhanced fluctuations close to the substrate, all occur when the liquid is near to bulk gas-liquid phase coexistence, i.e.\ when $\Delta\mu$ is small. Note that liquid water at ambient conditions is near to saturation. For water at ambient conditions $\beta\Delta\mu\sim10^{-3}$. This dimensionless quantity provides a natural measure of over-saturation, indicating where our results might be appropriate to water and to other solvents. The other key ingredient in determining the physics of effective interactions is the liquid-gas surface tension $\gamma_{lg}$, which is especially large for water. More precisely, it is the ratio $\gamma_{lg}/\Delta\mu\rho_l$, where $\rho_l$ is the density of the coexisting liquid, that sets the length scale for the capillary evaporation of any liquid; see Eq.~\eqref{eq:Kelvin's} below. The length scale in water is, of course, especially important. The influential article by Lum {\it et al.}~\cite{lum1999hydrophobicity} underestimates this. Subsequent articles~\cite{evans2004nonanalytic} and the informative piece~\cite{cerdeirina2011evaporation} by Cerdeiri{\~n}a {\it et al.\ }point to the fact that for water under ambient conditions the characteristic length for capillary evaporation is $L_c \sim \SI{1.5}{\micro\meter}$. The latter authors analyse why this length scale is so long and conclude this is due primarily to the large value of $\gamma_{lg}$ of water at room temperature. 

The paper is arranged as follows: In Sec.~\ref{sec:Method} we define the model solvent and the DFT used to describe it. Results for the fluid at a single planar substrate (wall) and between two identical walls are discussed in Sec.~\ref{sec:PlanarWall}. Then, in Sec.~\ref{sec:TwoBlocks}, we build a model for the two rectangular blocks and analyse the density profiles and local compressibility around the pair of blocks, comparing with results for the planar substrates. We examine the effect of changing the distance between the two blocks; this enables us to determine effective solvent mediated interactions between the blocks. These interactions differ enormously between an identical pair of solvophobic and a pair of solvophilic blocks. We also consider the case of i) a solvophobic and a solvophilic block and ii) blocks made from up to three patches that can be either solvophobic or solvophilic. The density and local compressibility profiles exhibit a rich structure in these cases and the resulting effective interactions exhibit considerable variety. We conclude in Sec.~\ref{sec:Conclusions} with a discussion of our results.

\section{Model solvent}
\label{sec:Method}

DFT\cite{evans1979nature,evans1992density,hansen2013theory} introduces the thermodynamic grand potential functional $\Omega[\rho]$ as a functional of the fluid one-body density profile, $\rho(\vec{r})$. The profile which minimises $\Omega[\rho]$ is the equilibrium profile and for this profile the functional is equal to the grand potential for the system. For a fluid of particles interacting via a hard-sphere pair potential plus an additional attractive pair potential $v(r)$, the grand potential functional can be approximated as follows:\cite{evans1979nature,evans1992density,hansen2013theory}
\begin{align}
	\Omega[\rho] &= \F_\mathrm{id}[\rho] + F_\mathrm{ex}^\mathrm{HS}[\rho] \nonumber\\ 
		&+ \frac{1}{2} \iint \rho(\vec{r}) \rho(\vec{r}')v(|\vec{r} - \vec{r}'|) \, \mathrm{d}\vec{r} \, \mathrm{d}\vec{r}'  \nonumber\\
		&+ \int \rho(\vec{r})(\phi(\vec{r}) - \mu ) \,  \mathrm{d}\vec{r} ,
		\label{eq:DFTmain}
\end{align}
where $\F_{\rm{id}} =  k_B T \int \rho(\vec{r}) ( \ln [ \Lambda^3\rho(\vec{r})] - 1 )\,\mathrm{d}\vec{r} $ is the ideal-gas contribution to the free energy, with Boltzmann's constant $k_B$, temperature $T$ and thermal de Broglie wavelength $\Lambda$. $\F_{\rm{ex}}^{\rm{HS}} = \int \Phi(\{n_\alpha\}) \,\mathrm{d}\vec{r}$ is the hard-sphere contribution to the excess free energy, which we treat using the White-Bear version of FMT,\cite{roth2002fundamental, roth2010fundamental} i.e.\ the free energy density $\Phi$ is a function of the weighted densities $\{n_\alpha\}$. $\phi(\vec{r})$ is the external potential and $\mu$ is the chemical potential. The attractive interaction between the particles is assumed to be given by a simple interaction potential, incorporating London dispersion forces,
\begin{equation}
    v(r) = \left\{
           \begin{array}{l l}
             -4\e \( \frac{\sigma}{r} \)^6
               & \quad  {r \ge \sigma}
             \\
             0
               & \quad {r < \sigma},
           \end{array}
         \right.
       \label{eq:attLJ}
\end{equation}
where $\sigma$ is the hard-sphere diameter and $\e>0$ is the attraction strength.

\begin{figure}[t]
	\includegraphics[width=0.495\textwidth]{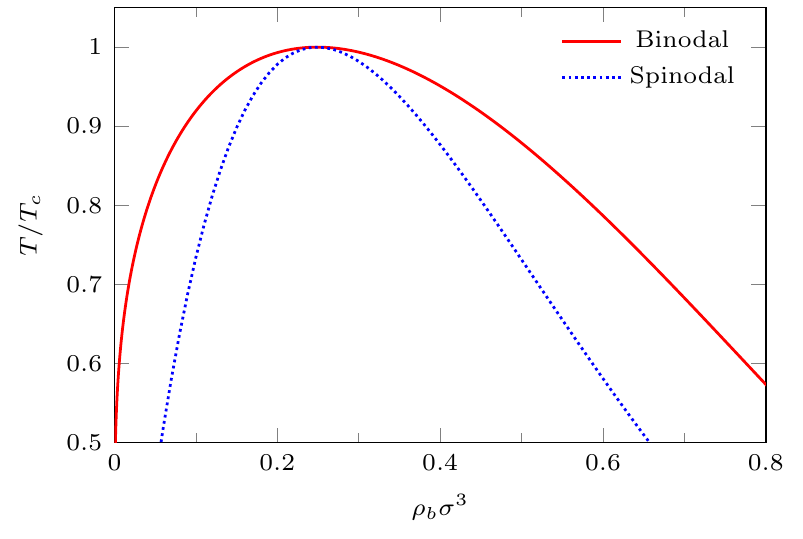}
	\caption{Bulk fluid phase diagram in the density versus temperature plane. $T_c$ is the critical temperature.}
	\label{fig:FMT:PhaseDiagram}
\end{figure}

\begin{figure}[t]
	\includegraphics[width=0.495\textwidth]{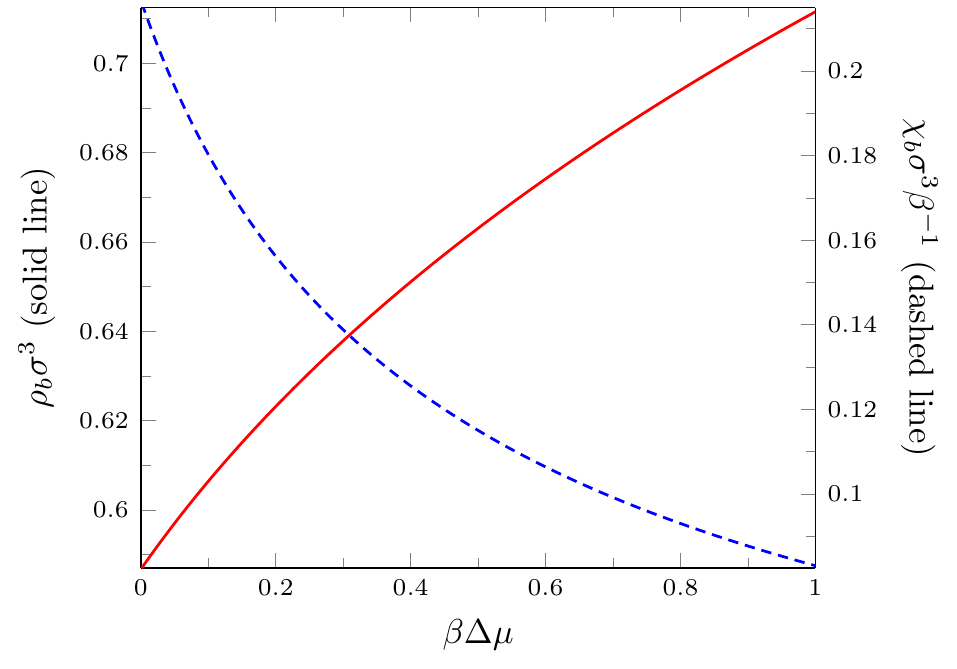}
	\caption{The bulk liquid density (solid line) and bulk compressibility (dashed line) as a function of $\Delta\mu=(\mu-\mu_{\rm coex})$, for fixed temperature $T=0.8\,T_c$.}
	\label{fig:FMTT80DMuRhobChib}
\end{figure}

In Fig.~\ref{fig:FMT:PhaseDiagram} we display the bulk fluid phase diagram, showing the gas-liquid coexistence curve (binodal) and spinodal calculated from Eq.~\eqref{eq:DFTmain}. Bulk gas-liquid phase separation occurs when $T < T_c$, where the critical temperature $T_c= 1.509\e/k_B$ and the critical density $\rho_c\sigma^3 = 0.249$. The results presented in the remainder of the paper are calculated along the isotherm with $T=0.8\,T_c$. We approach bulk gas-liquid coexistence from the liquid side, varying the chemical potential to determine the bulk liquid density. At coexistence, the chemical potential $\mu = \mu_\mathrm{coex}$ takes the same value for both liquid and gas phases. We define $\Delta\mu = \mu - \mu_\mathrm{coex}$, which gives a measure of how far a given bulk state is from coexistence. In Fig.~\ref{fig:FMTT80DMuRhobChib} we display the bulk liquid density as a function of $\beta\Delta\mu$, for $T = 0.8\,T_c$.

In addition to calculating density profiles and thermodynamic properties of the system, we also calculate the local (position dependent) compressibility in Eq.~\eqref{eq:chi}. In order to calculate this quantity, we use the finite difference approximation:
\begin{equation}
	\chi(\vec{r}) = \frac{\rho(\vec{r}; \mu + \delta \mu)-\rho(\vec{r}; \mu- \delta \mu)}{2\delta \mu},
\end{equation}
with $\beta\delta\mu=10^{-4}$. The bulk value of the compressibility $\chi_b\equiv(\partial\rho_b/\partial\mu)_T$, as a function of the chemical potential, is also shown in Fig.~\ref{fig:FMTT80DMuRhobChib}, for $T=0.8\,T_c$. We see that as the chemical potential is increased away from the value at coexistence, the bulk density increases (solid line) and $\chi_b$ decreases (dashed line).

\section{Liquid at planar walls}
\label{sec:PlanarWall}

Before presenting results for the liquid solvent around various different rectangular blocks, we describe its behaviour in the presence of a single planar wall and confined between two parallel planar walls. This is a prerequisite for understanding the behaviour around the blocks.

\subsection{Single hard wall with an attractive tail}

Initially, we treat the wall as being made of a different species of particle having a uniform density distribution and interacting with the fluid via a pair potential of the same form as the potential between the fluid particles, i.e.\ a hard-sphere potential together with the attractive pair potential
\begin{equation}
    v^h_{wf}(r) = \left\{
           \begin{array}{l l}
             -4\e_{wf}^{h} \( \frac{\sigma}{r} \)^6
               & \quad  {r \ge \sigma}
             \\
             0
               & \quad {r < \sigma}.
           \end{array}
         \right.
       \label{eq:WFH_pot}
\end{equation}
 This is the same as the potential in Eq.~\eqref{eq:attLJ}, but with $\e$ replaced by the wall-fluid attraction strength parameter $\e_{wf}^h>0$. Thus, the external one-body potential due to a substrate made of particles having uniform density $\rho_w$, occupying the half space $z<0$ (i.e.~the wall surface is located at $z = 0$), is
\begin{equation}
\phi(\vec{r})\equiv\phi(z)=\rho_w\int_{z<0}\mathrm{d}\vec{r}' v^h_{wf}(|\vec{r} - \vec{r}'|),
\label{eq:pot_int}
\end{equation}
for $z\geq\sigma/2$ and $\phi(z)=\infty$ for $z<\sigma/2$. From this we obtain
\begin{equation}
    \phi(z) = \left\{
           \begin{array}{l l}
             -\frac{2}{3}\e_{wf}^h\rho_w\sigma^3\pi \(\frac{\sigma}{z}\)^3
               & \quad  {z \ge \sigma/2}
             \\
             \infty
               & \quad {z < \sigma/2},
           \end{array}
         \right.
       \label{eq:FMT1D:EwH}
\end{equation}
where $z$ is the perpendicular distance from the surface of the wall.  Henceforth, for simplicity, we set $\rho_w\sigma^3=1$.

\begin{figure}[t]
	\includegraphics[width=0.495\textwidth]{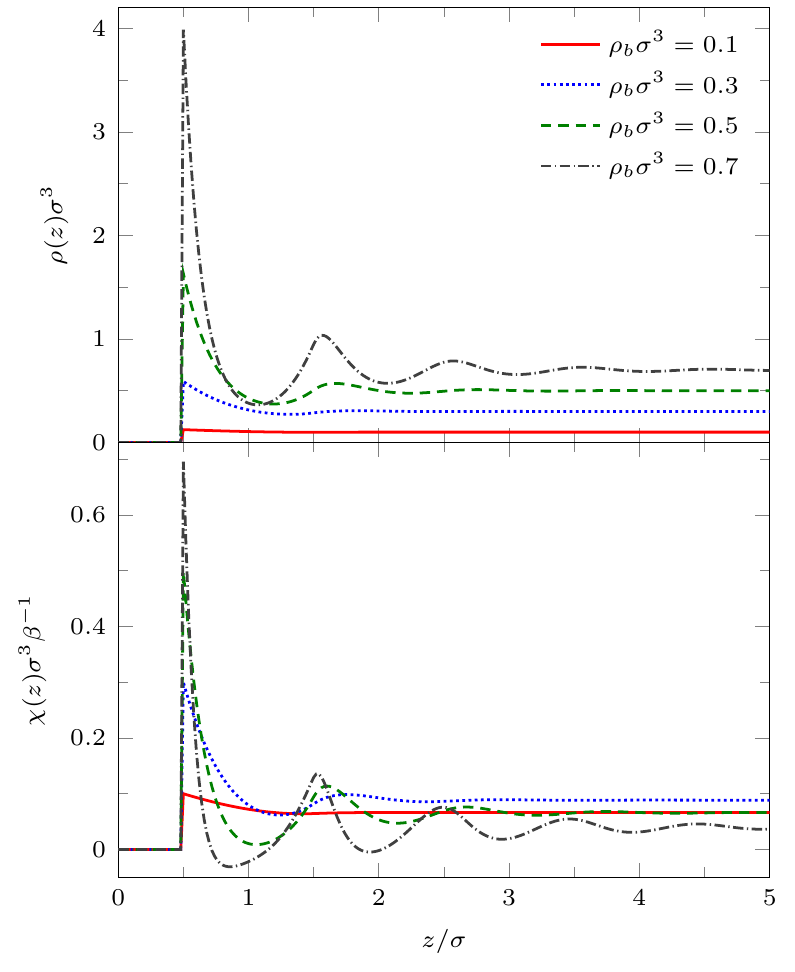}
	\caption{Density profiles $\rho(z)$ and local compressibility $\chi(z)$ for a fluid of hard spheres against a hard planar wall ($\e_{wf}^h = 0$) for different bulk densities.}
	\label{fig:FMT1D:E0Ew0}
\end{figure}

In Fig.~\ref{fig:FMT1D:E0Ew0} we display the fluid density profiles and the local compressibility for the hard-sphere fluid ($\e = 0$, equivalent to $T\to\infty$) against a planar hard wall ($\e_{wf}^h = 0$). This is useful for comparing with the later results, in order to assess the influence of the attractive interactions. We see that for low values of the bulk fluid density $\rho_b$, the density profile has little structure, as does $\chi(z)$. Increasing the bulk fluid density, we observe oscillations developing near to the wall arising from packing. The local compressibility $\chi(z)$ also develops significant oscillations near the wall. For higher values of $\rho_b$ we see that the contact value $\chi(\frac{\sigma}{2}^+)$ is significantly larger than the bulk value. We also note that it is possible for the local compressibility $\chi(z)$ to be {\em negative}, while of course the bulk value $\chi_b$ must be positive. This is because for larger values of $\rho_b$, the local density values at the minima of the oscillations are much smaller than in bulk, reflecting the fact that layering of the fluid at the wall becomes more pronounced.

\begin{figure}[t]
	\includegraphics[width=0.495\textwidth]{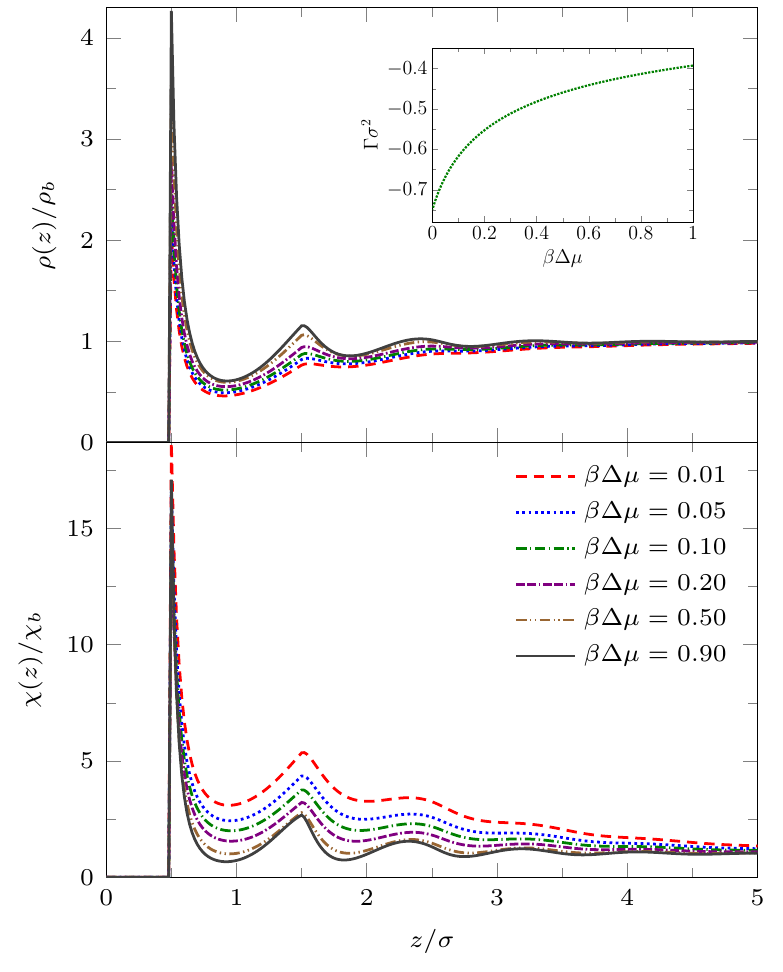}
	\caption{Scaled density profile and local compressibility for the liquid with temperature $T=0.8\,T_c$ at a single planar wall, Eq.~\eqref{eq:FMT1D:EwH}, with $\beta\e_{wf}^h = 0.13$. The corresponding bulk density $\rho_b$ and compressibility $\chi_b$ can be obtained from Fig.~\ref{fig:FMTT80DMuRhobChib}. Note that the contact angle for this choice of parameters is $\theta \approx 144^\circ$. {The inset in the upper panel shows the adsorption as a function of $\beta\Delta\mu$.}}
	\label{fig:FMT1DT80Ew013HS}
\end{figure}

We turn now to the case $\e>0$ and consider the temperature $T = 0.8\,T_c$, where bulk gas-liquid phase separation occurs. We set the wall attraction to be $\beta\e_{wf}^h = 0.13$, which is rather weak, corresponding to a solvophobic substrate with contact angle $\approx 144^\circ$ at this temperature. The contact angle $\theta$ is calculated using Young's formula
\begin{equation}
	\gamma_{wg} = \gamma_{wl} + \gamma_{gl}\cos\theta,
	\label{eq:Young's}
\end{equation}
where $\gamma_{wg}$, $\gamma_{wl}$ and $\gamma_{gl}$ are the wall-gas, wall-liquid and gas-liquid surface tensions, respectively. These interfacial tensions are each calculated separately via DFT in the usual manner (see e.g.\ Ref.~\onlinecite{stewart2012phase} and references therein).

Fig.~\ref{fig:FMT1DT80Ew013HS} shows liquid density profiles and the local compressibility (both divided by their respective bulk values) on the isotherm $T = 0.8\,T_c$. At this temperature the bulk density of the liquid at coexistence with the gas is $\rho_b\sigma^3 \approx 0.587$. For larger values of $\beta\Delta\mu$, away from coexistence, the density profiles exhibit oscillations at the wall, similar to the density profile for pure hard-spheres against the hard wall (Fig.~\ref{fig:FMT1D:E0Ew0}). As coexistence is approached, the oscillations in the density profiles are slightly eroded, although for this value of $\beta\e_{wf}^h = 0.13$, the changes in the density profile are not particularly striking. {This can also be seen from the inset in Fig.~\ref{fig:FMT1DT80Ew013HS} which displays the adsorption
\begin{equation}\label{eq:adsorption}
\Gamma=\int_0^\infty \mathrm{d}z (\rho(z)-\rho_b).
\end{equation}
Note that $\Gamma$ is negative and remains finite as $\beta\Delta\mu\to0$.} However, as can be seen from the lower panel of Fig.~\ref{fig:FMT1DT80Ew013HS}, where we display the corresponding local compressibility profiles $\chi(z)$, there is a significant increase in the local compressibility in layers adjacent to wall as coexistence is approached, $\beta\Delta\mu\to0^+$. We note also that the compressibility has oscillations whose maxima match those in the density profiles.

\subsection{Single soft Lennard-Jones wall}

\begin{figure}[t]
	\includegraphics[width=0.495\textwidth]{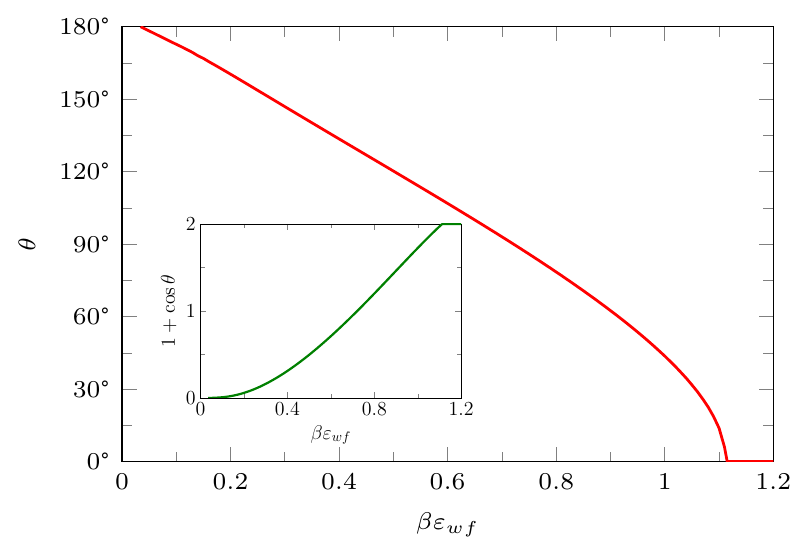}
	\caption{Contact angle $\theta$ as a function of the wall attraction strength $\beta\e_{wf}$ in Eq.~\eqref{eq:FMT1D:Ew} for $T=0.8\,T_c$. The inset plots $(1+\cos\theta)$ versus $\beta\e_{wf}$. Note that $(1+\cos\theta)$ approaches zero tangentially at $\beta\e_{wf}=0.0344$. By contrast, $(1+\cos\theta)$ approaches 2 linearly at $\beta\e_{wf}=1.11$. Thus, drying is critical and wetting is first order for this choice of wall.}
  \label{fig:FMT1DT80ThetaEw}
\end{figure}

The wall, Eq.~\eqref{eq:FMT1D:EwH}, considered in the previous subsection leads to the fluid density profile and local compressibility having a very sharp (and discontinuous) first peak at $z=\sigma/2$, particular to this wall potential. {The contact density $\rho(\sigma/2)$ is related to the bulk pressure via a sum rule [see e.g.~Eq~(68a) in Ref.~\onlinecite{henderson1992}], which is satisfied by the present DFT. For general wall-potentials there is no {\em explicit} formula for $\rho(\sigma/2)$. However, it is clear from the relation emerging from the sum rule that this quantity must be very large for a potential such as \eqref{eq:FMT1D:EwH}.\cite{henderson1992}} Real molecular fluids interact with substrates via continuous (softer) potentials. Thus, we now consider a planar wall composed of particles that interact with the fluid particles via the LJ pair potential
\begin{equation}
    v_{wf}(r) = 4\e_{wf} \left[  \( \frac{\sigma}{r} \)^{12} - \( \frac{\sigma}{r} \)^6 \right],
       \label{eq:WF_pot}
\end{equation}
where $\e_{wf}>0$ is the coefficient determining the strength of wall attraction. Thus, using Eq.~\eqref{eq:pot_int} with $v_{wf}^h$ replaced by $v_{wf}$, for $z>0$ and $\phi(z)=\infty$ for $z \le 0$, we have
\begin{equation}
    \phi(z) = \left\{
           \begin{array}{l l}
             4\e_{wf}\rho_w\sigma^3\pi \left( \dfrac{\sigma^{9}}{45z^9} - \dfrac{\sigma^3}{6z^3}\right)
               & \quad  {z > 0}
             \\
             \infty
               & \quad {z \le 0},
           \end{array}
         \right.
       \label{eq:FMT1D:Ew}
\end{equation}
where $z$ is the perpendicular distance from the wall. The contact angle calculated using Eq.~\eqref{eq:Young's} for the liquid against this soft wall for $T=0.8\,T_c$ is shown in Fig.~\ref{fig:FMT1DT80ThetaEw}. When we set the wall attraction to be $\beta\e_{wf} = 0.3$, then the contact angle is $\theta \approx 144^\circ$, which is the same contact angle that the fluid has against the hard wall with an attractive tail potential \eqref{eq:FMT1D:EwH}, with $\beta\e^h_{wf} = 0.13$ -- as treated in Fig.~\ref{fig:FMT1D:E0Ew0}. Note that in Fig.~\ref{fig:FMT1DT80ThetaEw} the drying transition, where $\theta\to 180^\circ$, occurs at $\beta\e_{wf}=0.0344$ and is continuous (critical). The numerical result from DFT for this value agrees precisely with the analytical prediction from the binding potential treatment for the same model potentials treated in the sharp-kink approximation.\cite{stewart2005critical} The latter predicts a continuous drying transition when $\beta\e_{wf}(\rho_w\sigma^3) = \beta\e(\rho_g\sigma^3)$, where $\rho_g$ is the density of the coexisting gas at the given temperature. What is striking about this result is that it also applies for the wall potential in Eq.~\eqref{eq:FMT1D:EwH}, i.e.\ critical drying occurs at the same value $\beta\e_{wf}^h=\beta\e_{wf}=0.0344$. This is a consequence of both potentials having the same asymptotic decay as $z\to\infty$. However, for the potential in Eq.~\eqref{eq:FMT1D:EwH} wetting, $\theta=0$, occurs at a much smaller value of $\beta\e_{wf}^h$. Thus, the overall behaviour of $(1+\cos\theta)$ vs wall strength is sensitive to the precise form of the wall potential.

\begin{figure}[t]
	\includegraphics[width=0.495\textwidth]{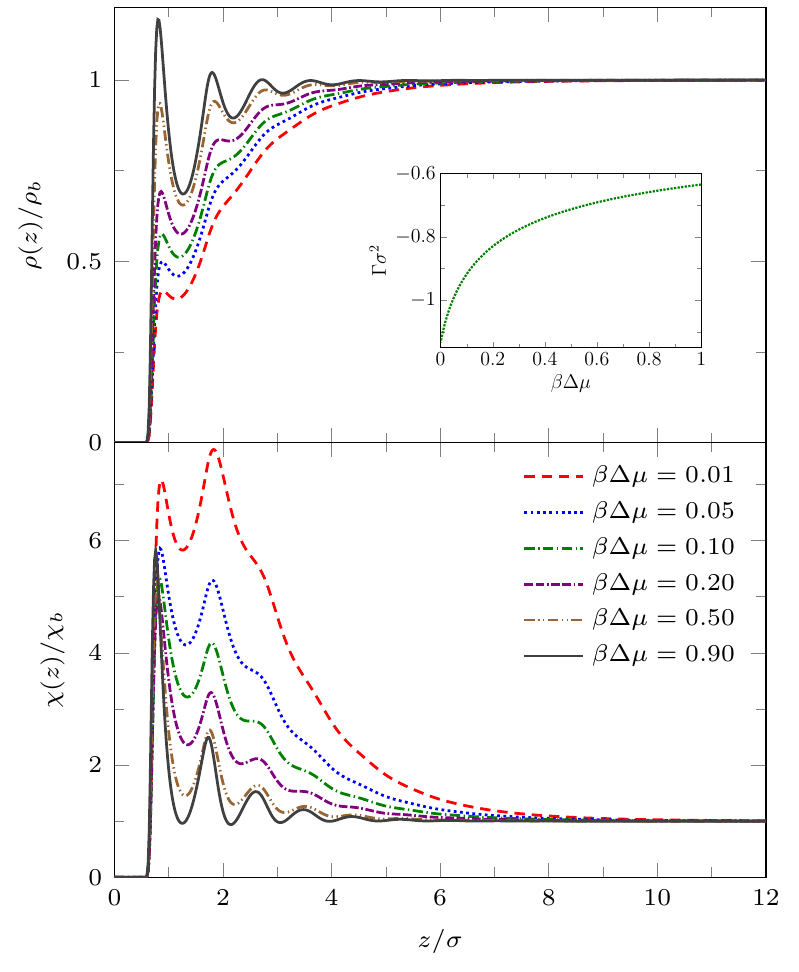}
	\caption{ Scaled liquid density profile and local compressibility at a single planar wall, Eq.~\eqref{eq:FMT1D:Ew}, with $T=0.8\,T_c$ and $\beta\e_{wf} = 0.3$. The bulk densities corresponding to the chemical potentials $\beta\Delta\mu$ in the key can be read from Fig.~\ref{fig:FMTT80DMuRhobChib}. Note that the contact angle for this choice of parameters is $\theta \approx 144^\circ$. {The inset in the upper panel shows the adsorption as a function of $\beta\Delta\mu$.}}
  \label{fig:FMT1DT80Ew03}
\end{figure}

Fig.~\ref{fig:FMT1DT80Ew03} shows liquid density profiles and the local compressibility (both divided by their respective bulk values) on the isotherm $T = 0.8\,T_c$. For large values of $\beta\Delta\mu$ the density profiles exhibit oscillations at the wall similar to the density profiles for the walls in Figs.~\ref{fig:FMT1D:E0Ew0} and \ref{fig:FMT1DT80Ew013HS}. However, as coexistence is approached the degree to which the oscillations in the density profiles near the wall are eroded is greater than for the case of the wall~\eqref{eq:FMT1D:EwH} and a region of depleted density appears at the wall. Note that for this value of $\e_{wf}$, the low density film close to the wall remains finite in thickness right up to coexistence, $\beta\Delta\mu\to0$, since the wall-liquid interface is only partially dry: $\theta<180^\circ$. {This can also be seen from the inset which shows the adsorption \eqref{eq:adsorption}. Although $\Gamma$ is somewhat larger in magnitude than for the wall potential \eqref{eq:FMT1D:EwH}, displayed in the inset to Fig.~\ref{fig:FMT1DT80Ew013HS}, it remains finite at coexistence.} In the lower panel of Fig.~\ref{fig:FMT1DT80Ew03} we display the corresponding local compressibility profiles $\chi(z)$ in the vicinity of this solvophobic surface. We observe that in the first few adsorbed layers, the local compressibility increases significantly i.e.\ the range over which $\chi(z)/\chi_b$ is significantly greater than unity increases as $\beta\Delta\mu\to0$. Moreover the maximum near $z/\sigma=2$, corresponding to the second particle layer, grows rapidly as $\beta\Delta\mu$ decreases.

\subsection{Two planar walls}

\begin{figure}[t]
	\includegraphics[width=0.495\textwidth]{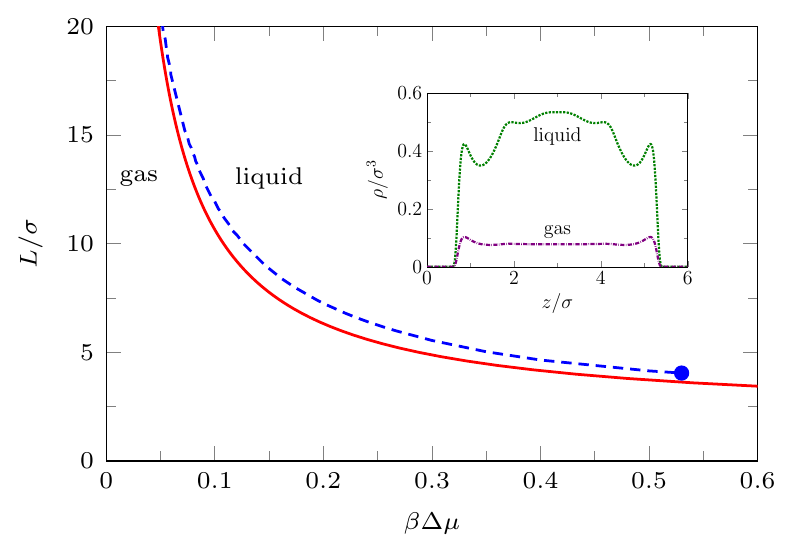}
	\caption{The figure shows the capillary evaporation line calculated via DFT (dashed) with that from the Kelvin equation \eqref{eq:Kelvin's} (solid), for two parallel planar walls with $\beta\e_{wf}=0.3$ and $T=0.8\,T_c$. The critical point of the capillary evaporation is marked with a circle. For values of $L$ below the critical value, there is no capillary evaporation. The inset shows the coexisting gas and liquid density profiles when $L=6\sigma$, i.e.\ $\beta\Delta\mu \approx 0.266$. }
  \label{fig:FMT1D2T80Ew03LDMu}
\end{figure}

We now consider briefly a pair of planar walls, where the distance between the walls is $L$. The external potential is
\begin{equation}
\phi_{2w}(z)=\phi(z)+\phi(L-z),
\end{equation}
where $\phi(z)$ is given by the soft wall Eq.~\eqref{eq:FMT1D:Ew}. Capillary evaporation from this planar slit can occur as $\beta\Delta\mu\to0$, whereby the liquid between the two solvophobic planar walls evaporates as coexistence is approached. The value of $L$ at which this occurs can be estimated from the Kelvin equation:\cite{evans1987phase, tarazona1987phase}
\begin{equation}
	L^* \approx \frac{-2\gamma_{lg}\cos\theta}{\Delta\mu(\rho_l-\rho_g)}
	\label{eq:Kelvin's}
\end{equation}
where $L^* \equiv L-2\sigma$ is defined as roughly the distance between maxima of the density profile, corresponding to the first adsorbed layer at each wall. $L^*$ is the effective distance between the walls. $\gamma_{lg}$ is the gas-liquid interfacial tension, $\theta$ is the single planar wall contact angle, and $\rho_g$ and $\rho_l$ are the coexisting gas and liquid densities, respectively. Eq.~\eqref{eq:Kelvin's} is appropriate to a partial drying situation.\cite{evans1987phase}

Fig.~\ref{fig:FMT1D2T80Ew03LDMu} shows the capillary evaporation phase transition line, comparing the prediction from the Kelvin equation \eqref{eq:Kelvin's} with that calculated from DFT. This is the line in the $(\Delta\mu,L)$ plane where the gas-filled slit and liquid-filled slit are at thermodynamic coexistence, i.e.\ these states have the same grand potential, temperature, and chemical potential. The inset in Fig.~\ref{fig:FMT1D2T80Ew03LDMu} shows the density profiles of coexisting gas and liquid states for $L=6\sigma$. As we expect, the Kelvin equation is accurate for large $L$, but is less reliable for small $L$. Nevertheless for values down to $L\approx4\sigma$ and $\beta\Delta\mu=0.53$, where the critical point occurs in DFT, the Kelvin equation prediction remains fairly good. This may come as a surprise to some readers, given that the equation is based on macroscopic thermodynamics. Note that Eq.~\eqref{eq:Kelvin's} does not account for a capillary critical point.\cite{evans1987phase,tarazona1987phase} We have also investigated the solvent mediated potential between two planar walls, i.e.\ the excess grand potential arising from confinement. The derivative of this quantity with respect to $L$ jumps at capillary evaporation. We return to this later.

\section{Two rectangular blocks}
\label{sec:TwoBlocks}

\begin{figure}[t]
	\includegraphics[width=0.47\textwidth]{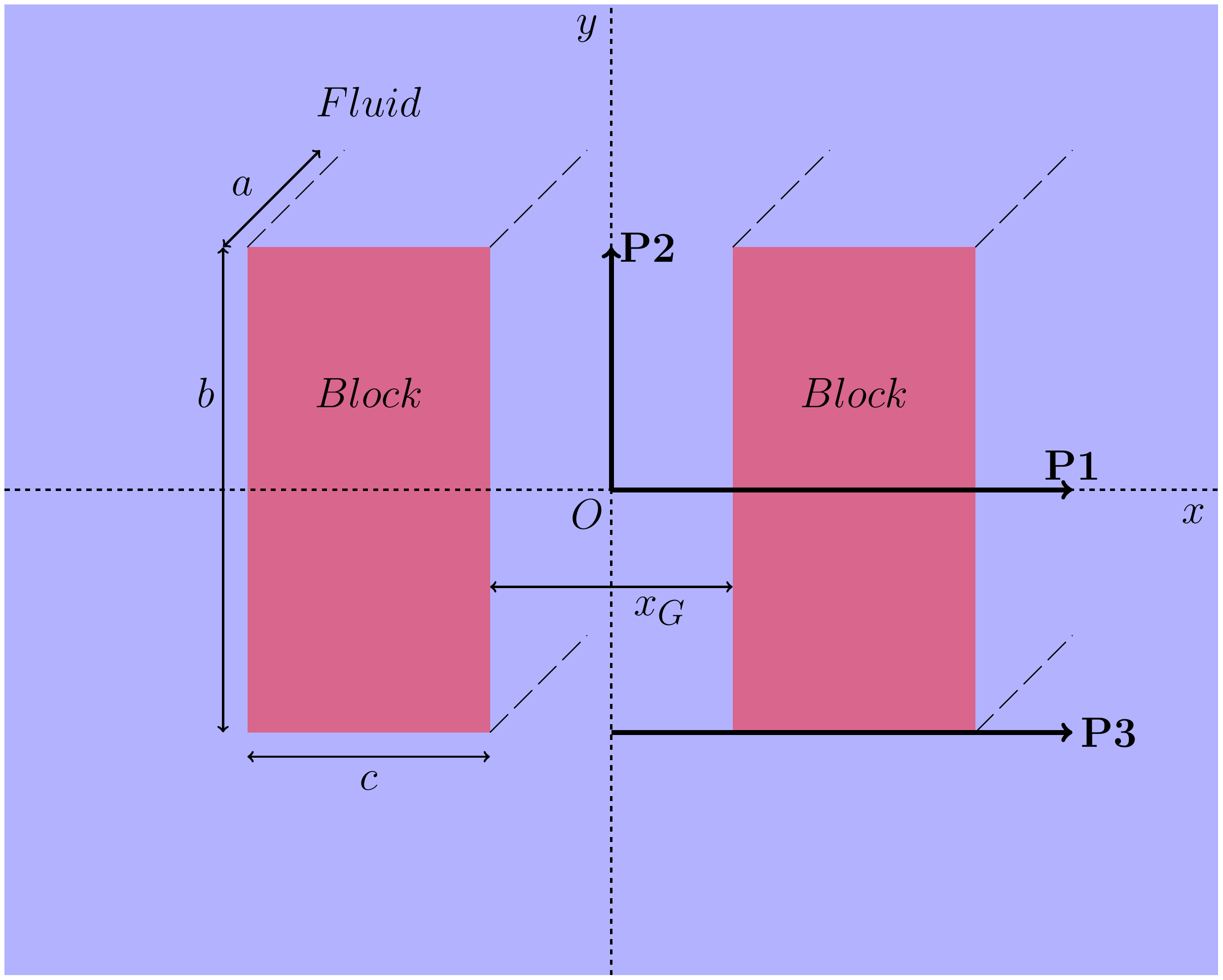}
	\caption{An illustration of two rectangular cross-section solid blocks immersed in the liquid. The cross sectional area of each block is $b\times c$ and the length of the blocks is $a$. We assume $a\to\infty$. In the case sketched here, the blocks are made of a uniform density of particles of the same diameter $\sigma$ as the liquid particles, interacting with the liquid particles via the potential in Eq.~\eqref{eq:WF_pot}. ``P1", ``P2" and ``P3" denote three different paths along which we display density profiles and the local compressibility in Figs.~\ref{fig:FMT2D:PM3D_Rho} to \ref{fig:FMT2D:E3:Ew05:centre}, below.}
	\label{fig:TwoBlock}
\end{figure}

In this section we describe the properties of the liquid around two rectangular cross-section beams of length $a$ -- the two ``blocks" illustrated in Fig.~\ref{fig:TwoBlock}. We assume that the blocks are long, i.e.\ we take the limit $a\to\infty$. The distance between the closest faces of the blocks is $x_G$ and we set the size of the cross-section of the two blocks to be $b\times c$, where $b=8\sigma$ and $c=3\sigma$. We locate the origin of our Cartesian coordinate system to be midway between the two blocks.

The external potential due to the two blocks is defined in a manner analogous to that used above for the planar wall potential [cf.\ Eq.~\eqref{eq:pot_int}]; i.e.\ the potential due to the two blocks is
\begin{equation}
  \phi(\vec{r})=\rho_w\int_{{\cal D}}\mathrm{d}\vec{r}' v_{wf}(|\vec{r} - \vec{r}'|),
  \label{eq:pot_int_blocks}
\end{equation}
where ${\cal D}$ is the region of space occupied by each of the blocks. The parameter $\e_{wf}$ characterises the strength of the attraction between the blocks and the fluid. When $\e_{wf}$ is small, the blocks are solvophobic, but for larger values of $\e_{wf}$ they are solvophilic. Later we consider blocks having some sections that are solvophobic and others that are solvophilic: these are the so called ``patchy'' blocks. Note that the region ${\cal D}$ is where the fluid is completely excluded, with $\phi(\vec{r})=\infty$ and is made of two volumes with cross sectional area $b\times c=8\sigma\times3\sigma$. However, the effective exclusion cross-sectional area of each block is $\approx b^*\times c^* = 10\sigma\times5\sigma$, which includes an exclusion zone of width $\sigma$ around each of the blocks.

\subsection{Two solvophobic blocks}

\subsubsection{Blocks at fixed separation $x_G$}

\begin{figure*}[t]
	\includegraphics[width=0.9\textwidth]{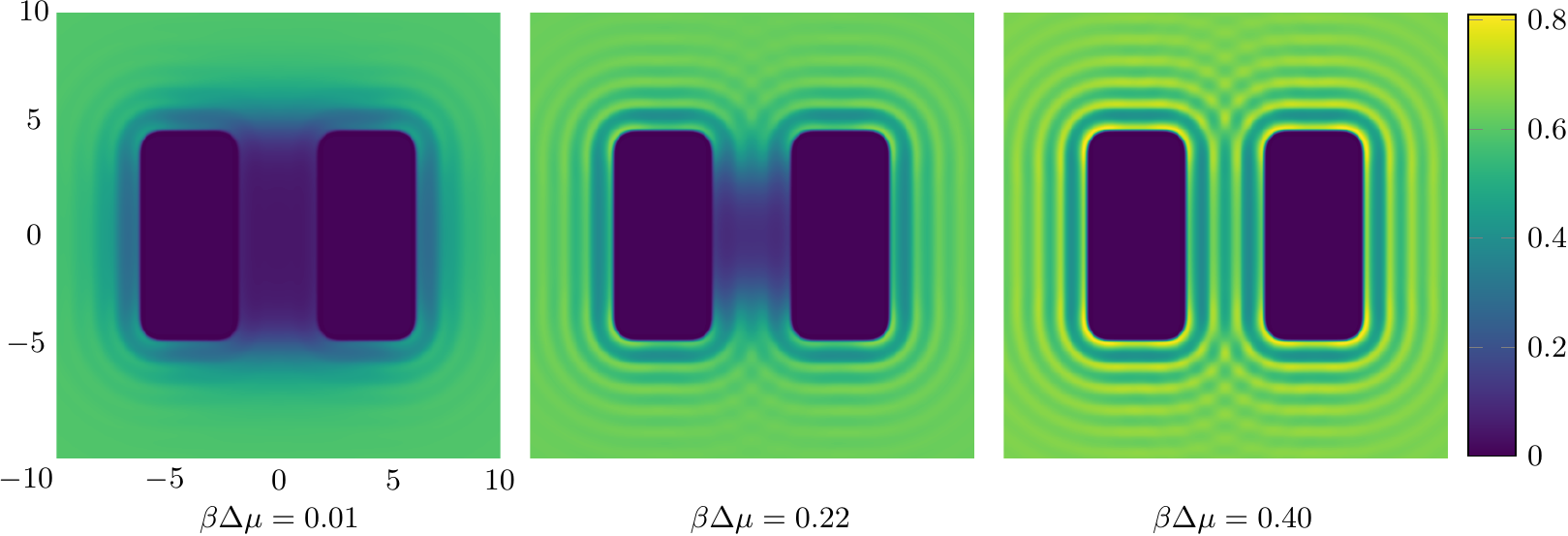}
	\caption{Density profile $\sigma^3\rho(x,y)$ around the pair of solvophobic blocks separated a distance $x_G = 5\sigma$ apart, for three values of the chemical potential. The temperature $T=0.8\,T_c$ and wall attraction strength $\beta\e_{wf} = 0.3$.}
	\label{fig:FMT2D:PM3D_Rho}
\end{figure*}

\begin{figure*}[t]
	\includegraphics[width=0.9\textwidth]{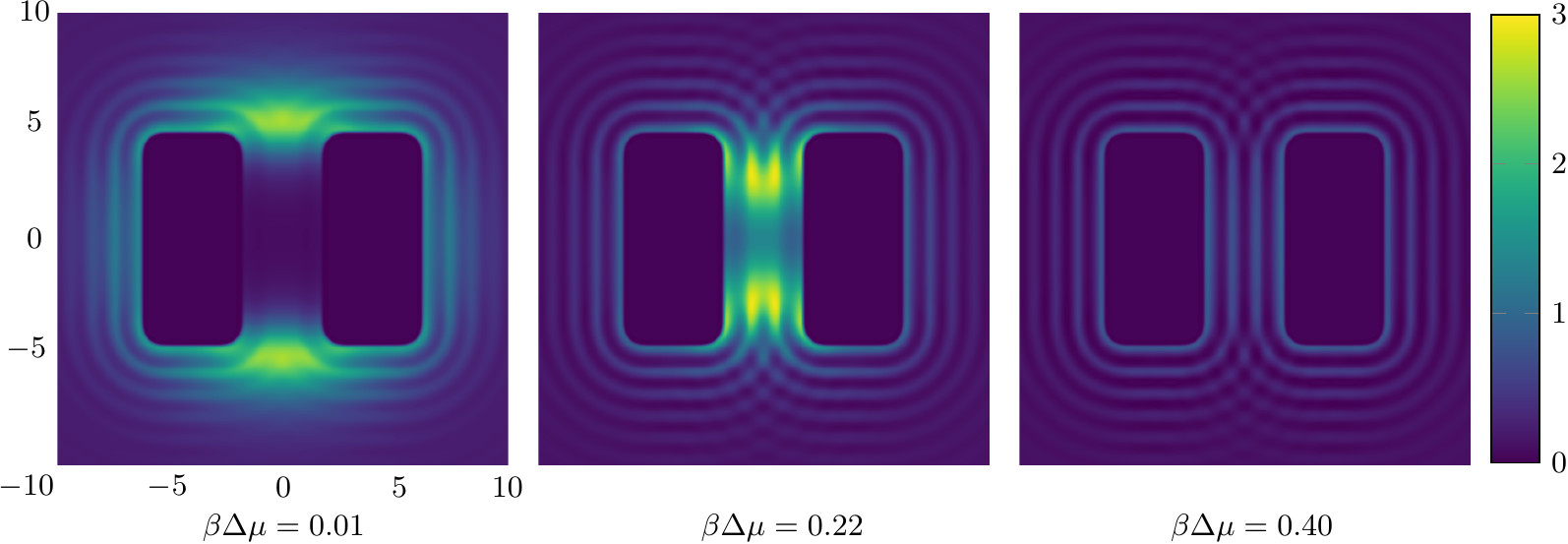}
	\caption{The local compressibility $k_BT\sigma^3\chi(x,y)$ around the pair of solvophobic blocks, for three states approaching bulk coexistence. The corresponding density profiles are displayed in Fig.~\ref{fig:FMT2D:PM3D_Rho}.}
	\label{fig:FMT2D:PM3D_Chi}
\end{figure*}

The results we present first are for a pair of blocks with soft solvophobic surfaces with $\beta\e_{wf}=0.3$, at the temperature $T = 0.8\,T_c$. Recall that for the single soft planar wall this value of $\e_{wf}$ corresponds to a contact angle $\theta \approx 144^\circ$ and that for the pair of planar walls the capillary evaporation critical point is at $\beta\Delta\mu=\beta\Delta\mu_{cc}=0.53$ -- see Fig.\ \ref{fig:FMT1D2T80Ew03LDMu}. In Figs.~\ref{fig:FMT2D:PM3D_Rho} and \ref{fig:FMT2D:PM3D_Chi} we display density profiles and the local compressibility $\chi(\vec{r})$, for various $\beta\Delta\mu$ and fixed $x_G=5\sigma$.

The density profiles in Fig.~\ref{fig:FMT2D:PM3D_Rho} show that as coexistence is approached, i.e.\ as $\beta\Delta\mu \to 0$, the density in the space between the pair of blocks becomes very small, i.e.\ gas-like. This is somewhat analogous to the capillary evaporation observed between two infinite planar walls -- see Fig.~\ref{fig:FMT1D2T80Ew03LDMu}. For larger values of $\beta\Delta\mu$, away from the value where bulk gas-liquid coexistence occurs, we see oscillations in the density profile arising from the packing of the liquid particles around the blocks. We also note that the density is higher near the corners of the blocks.

The local compressibility $\chi(\vec{r})$ provides a measure of the strength of the local fluctuations within the fluid and so large values of this quantity reveals regions in space where the local density fluctuations are greatest. In Fig.~\ref{fig:FMT2D:PM3D_Chi}, we see that for $\beta\Delta\mu = 0.4$, well away from bulk coexistence, the local compressibility is largest around the surface of the two blocks, decreasing in an oscillatory manner as the distance from the blocks increases. When the chemical potential deviation is smaller, $\beta\Delta\mu = 0.22$, the local compressibility in the vicinity of the outside of the blocks is similar to the case for the larger value of $\beta\Delta\mu = 0.4$. However, in the region between the two blocks, we see that the local compressibility is significantly larger, indicating strong fluctuations in this region. For $\beta\Delta\mu = 0.22$, we see from Fig.~\ref{fig:FMT2D:PM3D_Rho} that the average density in the gap between the blocks is intermediate between the bulk gas and liquid coexisting densities and so we expect that typical microstates of the system include both gas-like and liquid-like average densities in the gap. The fluctuations of the system between these two typical states are what lead to the high values of the local compressibility.

Approaching even closer to the bulk coexistence point leads to the gas being stabilised in the gap between the blocks -- see the density profiles for $\beta\Delta\mu = 0.01$ in Fig.~\ref{fig:FMT2D:PM3D_Rho}. For this value of $\beta\Delta\mu$ we see from Fig.~\ref{fig:FMT2D:PM3D_Chi} that the region where the local compressibility is largest is not in the gap between the blocks, but is instead at the entrance to this region, where there is an `interface' between the bulk liquid and the gas-like intrusion between the blocks. It is the fluctuations in this interface that lead to the maxima in the local compressibility $\chi(x,y)$.

We now present results for $x_G = 7\sigma$, i.e.\ with the gap between the blocks being slightly larger. In order to display in more detail the properties of the density profiles and local compressibility around the pair of blocks, we plot these along the three different paths P1, P2 and P3, illustrated in Fig.~\ref{fig:TwoBlock}. The density and compressibility profiles are, of course, symmetrical around the mid-line through the gap between the blocks, so we display profiles around the right hand block only. From Fig.~\ref{fig:TwoBlock} we see that paths P1 and P2 are along the lines of symmetry and path P3 is along the horizontal side of the block.

\begin{figure*}[t]
	\includegraphics[width=0.495\textwidth]{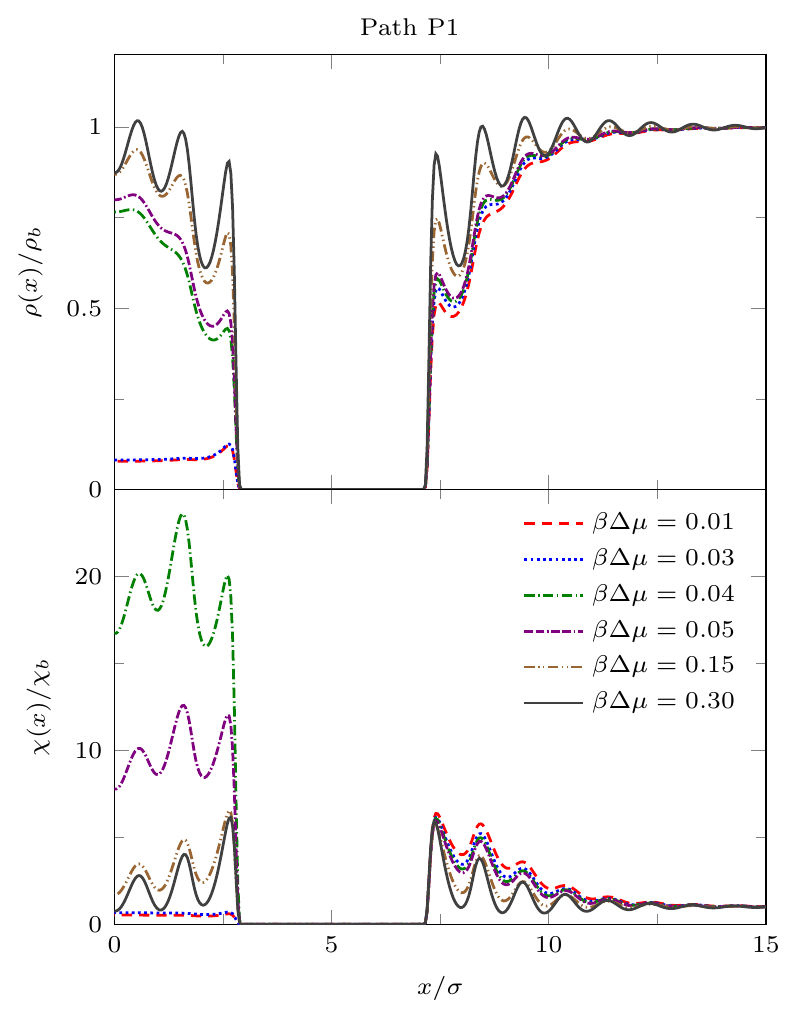}
	\includegraphics[width=0.495\textwidth]{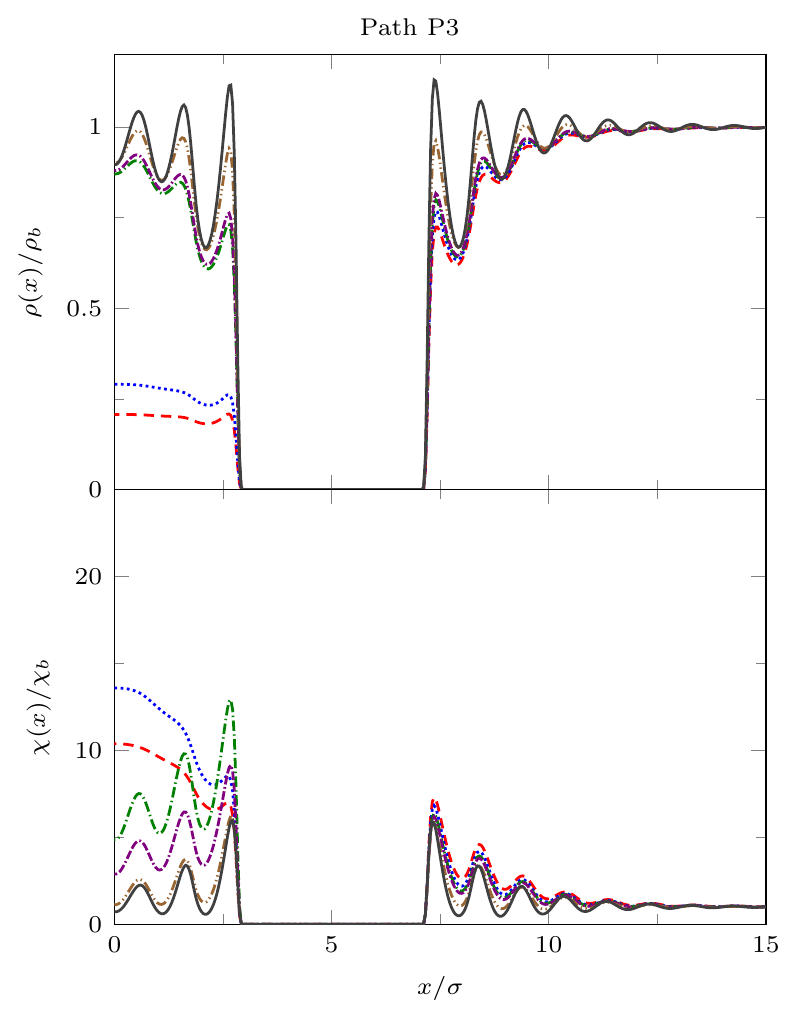}
	\caption{Left: density profiles $\rho(x) \equiv \rho(x,0)$ (top) and local compressibility $\chi(x) \equiv \chi(x,0)$ (bottom) of the fluid along path P1 for various values of the chemical potential and for $T=0.8\,T_c$, $\beta\e_{wf}=0.3$ and fixed distance $ x_G = 7\sigma$ between the blocks. Right: the corresponding functions $\rho(x) \equiv \rho(x,\pm b/2)$ and $\chi(x) \equiv \chi(x,\pm b/2)$ on path P3. Note that $\rho(\vec{r}), \chi(\vec{r}) \equiv 0$ within the block.}
	\label{fig:FMT2D:E3:Ew05:xGap7:P1nP3}
\end{figure*}

In Fig.~\ref{fig:FMT2D:E3:Ew05:xGap7:P1nP3}, we display results along paths P1 and P3. On both paths, both the density and the local compressibility are, of course, zero within the block. Focussing first along the portion of path P1 not in the gap between the blocks, we see that the profiles for varying $\beta\Delta\mu$ are very similar to those displayed in Fig.~\ref{fig:FMT1DT80Ew03} for the planar LJ wall: as $\beta\Delta\mu$ is decreased, the density in the vicinity of the wall decreases and the maxima in $\chi(\vec{r})$ near the wall increase. Comparing with the density profiles along the parallel path P3, along the horizontal size of the blocks, we see that away from the gap between the blocks the local density is slightly higher than along path P1 (this is the influence of the corner), but both the density and compressibility follow the same trend as along path P1.

Moving on to examine the behaviour in the gap between the blocks, in Fig.~\ref{fig:FMT2D:E3:Ew05:xGap7:P1nP3} we see that on decreasing $\beta\Delta\mu$, along path P1 the density decreases and at $\beta\Delta\mu \lesssim 0.04$ there is a discontinuous change in the density profile. The density profiles for $\beta\Delta\mu = 0.03$ and $0.01$ are almost identical and correspond to a dilute `gas' state. The strong fluctuations connected to the onset of this transition result in very large values of $\chi(x,0)$ for $\beta\Delta\mu = 0.05$ and $0.04$. $\chi(x,0)$ exhibits a discontinuous change in the gap between the blocks at the value of $\beta\Delta\mu$ where the density profile jumps. Moreover, along the portion of path P3 along the end of the gap between the blocks, we also observe a large jump in the density profile as coexistence is approached. Along path P3 the local compressibility also jumps. Unlike on path P1, where in the gap $\chi(x,0)$ takes small gas-like values for $\beta\Delta\mu=0.03$ and $0.01$, on path P3 $\chi(0,\pm b/2)/\chi_b$ is very large for these values of $\beta\Delta\mu$, reflecting the occurrence of gas-liquid interfacial fluctuations. All of this is reminiscent of the capillary evaporation observed for two planar solvophobic walls. However, the transition occurs at a smaller value of $\beta\Delta\mu$ due to the finite dimensions of the blocks. Specifically, the transition occurs at $\beta\Delta\mu \lesssim 0.04$, whereas for the planar slit with $L=7\sigma$ evaporation occurs at $\beta\Delta\mu=0.21$; see Fig.~\ref{fig:FMT1D2T80Ew03LDMu}.

\begin{figure}[t]
	\includegraphics[width=0.495\textwidth]{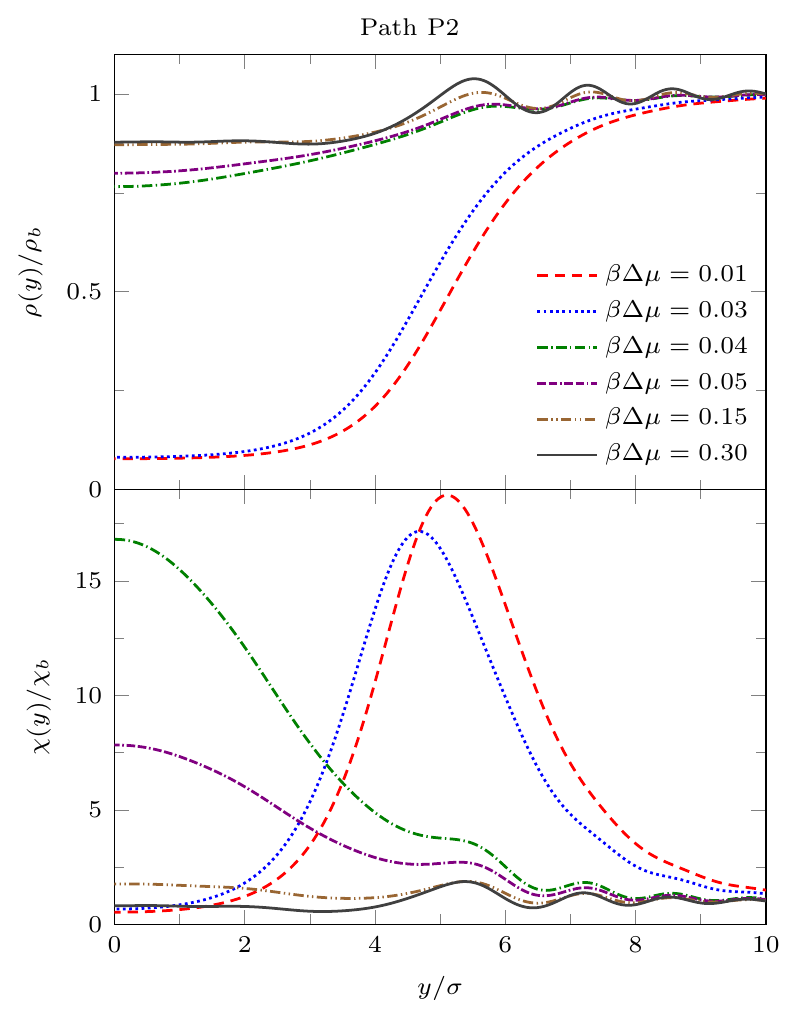}
	\caption{Density profiles  $\rho(y) \equiv \rho(0,y)$  and local compressibility  $\chi(y) \equiv \chi(0,y)$ along path P2 for various values of the chemical potential and for $T=0.8\,T_c$, $\beta\e_{wf}=0.3$ and fixed distance $ x_G = 7\sigma$ between the blocks. Path P2 goes from the mid point between the two blocks ($y=0$) into the bulk liquid ($y=\infty$) parallel to the vertical surfaces of the two blocks -- see Fig.~\ref{fig:TwoBlock}.}
	\label{fig:FMT2D:E3:Ew05:xGap7:P2}
\end{figure}

In Fig.~\ref{fig:FMT2D:E3:Ew05:xGap7:P2} we display density profiles and the local compressibility along path P2 (see Fig.~\ref{fig:TwoBlock}), which starts from the origin (the mid point between the blocks) and goes along the positive $y$-axis. For small $\beta\Delta\mu$, i.e.\ $\beta\Delta\mu = 0.03$ and $0.01$, we see that the density is gas-like in the gap between the blocks, increasing to the bulk liquid value outside the gap, $y \gtrsim 8\sigma$. The density profile changes discontinuously at $\beta\Delta\mu \lesssim 0.04$ and for larger values, the density is liquid-like throughout path P2. For smaller values of the chemical potential, $\beta\Delta\mu\lesssim0.04$, there is a local maximum in the local compressibility along this path and the location of the maximum occurs roughly where the density profile $\rho(0,y)/\rho_b = 0.5$. Thus, as the chemical potential is varied, the local compressibility maximum splits and shifts along the $y$-axis in the gap between the blocks. Recall that along the $y$-axis the system is symmetric around the origin, therefore for small $\beta\Delta\mu$ there is a peak in $\chi(\vec{r})$ at each of the entrances to the gap, i.e.\ for $y \approx \pm 5\sigma$ [cf.~Fig.~\ref{fig:FMT2D:PM3D_Chi}].

\subsubsection{Varying the separation between the blocks}
\label{subsec:vary}

\begin{figure}[t]
	\includegraphics[width=0.495\textwidth]{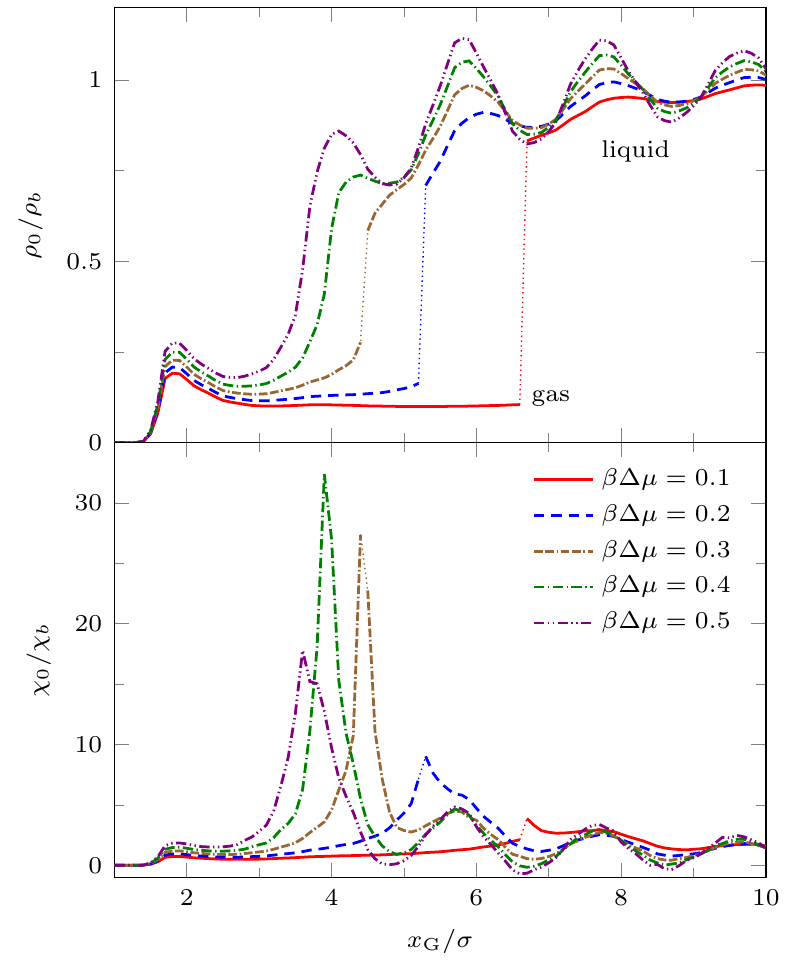}
	\caption{Mid point density, $\rho(0,0)=\rho_0$, and compressibility $\chi(0,0)=\chi_0$, as a function of the distance between the blocks $x_G$, for various values of the chemical potential. These are for fixed $T=0.8\,T_c$ and $\beta\e_{wf} = 0.3$. The jumps in $\rho_0$ are marked with dotted lines for $\beta\Delta\mu=0.1,0.2$ and $0.3$. There are accompanying jumps in $\chi_0$ at the same state points, that are not easy to ascertain on the scale of these plots. For $\beta\Delta\mu=0.5, 0.4$ we observe continuous variation of $\rho_0$ and $\chi_0$. There is no sharp, first order transition. A critical transition must occur, within mean-field, between $\beta\Delta\mu=0.3$ and $0.4$, resulting in a divergence of $\chi_0$.}
	\label{fig:FMT2D:E3:Ew05:centre}
\end{figure}

In Fig.~\ref{fig:FMT2D:E3:Ew05:centre} we show how the mid-point density $\rho(0,0) \equiv \rho_0$, varies as the distance between the two blocks $x_G$ is changed. The figure also shows how the local compressibility at the origin $\chi(0,0) \equiv \chi_0$ varies with $x_G$. For $\beta\Delta\mu=0.1,0.2 \text{ and }0.3$ there is a discontinuous change in the density. The magnitude of the `jump' gets larger as $\beta\Delta\mu$ approaches zero. Note that if the density, or more precisely the adsorption, jumps from one finite to another finite value at a particular value of $x_G$ then so must the local compressibility. This is a signature of the first order transition which occurs in the present mean-field DFT treatment. For $\beta\Delta\mu\gtrsim 0.4$ the density varies smoothly with $x_G$. In the lower panel of Fig.~\ref{fig:FMT2D:E3:Ew05:centre} we observe a peak in $\chi_0$ when the mid-point density crosses $\rho_0/\rho_b = 0.5$. The height of this peak appears to be maximal at $\beta\Delta\mu\approx 0.4$, the value at which the transition in $\rho_0$ appears to change from discontinuous to continuous. In other words, capillary evaporation still manifests itself as a first order transition, with its accompanying critical point, in our mean-field treatment of `evaporation' between two blocks of finite cross-sectional area. Bearing in mind the effectively one-dimensional nature ($b$ and $c$ finite but $a \to \infty$) of the capillary-evaporation-like transition we observe in the fluid between the blocks, we expect the divergence in $\chi_0$ to be rounded, in reality, by finite size effects. Likewise, we expect the jump in $\rho_0$ to be rounded in reality.

\begin{figure}[t]
	\includegraphics[width=0.49\textwidth]{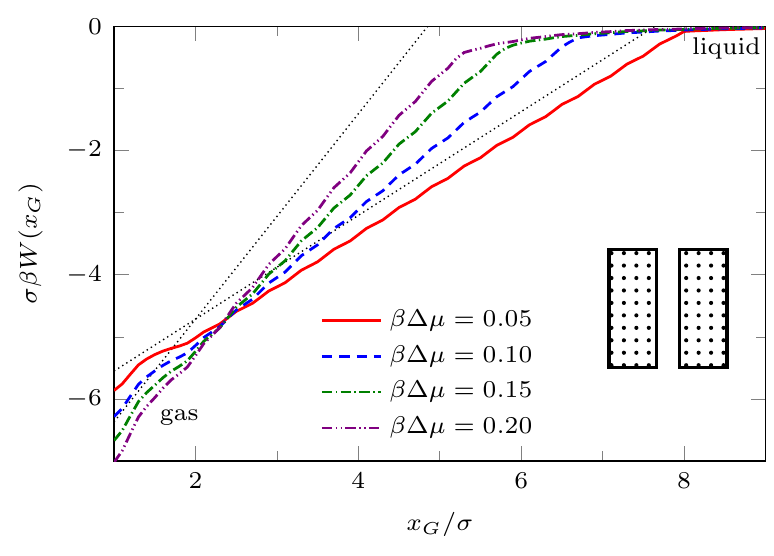}
	\caption{The solvent mediated potential (excess grand potential) between a pair of solvophobic blocks, as a function of distance between blocks $x_G$ for fixed $T=0.8\,T_c$ and $\beta\e_{wf} = 0.3$. The dotted lines are the estimates for the two cases $\beta\Delta\mu=0.05$ (lower) and $0.20$ (upper) calculated from Eq.~\eqref{eq:SolventMediatedPotentialTwoBlock}, with $E=0$. The DFT results display two branches -- see text.}
	\label{fig:FMT2DT80Omega_xGapS1E0}
\end{figure} 

In Fig.~\ref{fig:FMT2DT80Omega_xGapS1E0} we display a plot of the excess grand potential per unit length, $W(x_G)\equiv(\Omega(x_G)-\Omega_\infty)/a$, as a function of $x_G$. $\Omega_\infty \equiv \Omega(x_G\to\infty)$ is the value of the grand potential when the two blocks are far apart. $W(x_G)$ is the solvent mediated interaction potential per unit length between the two blocks. Since $W(x_G)$ becomes increasingly negative as $x_G$ decreases, this indicates that the solvent mediated interaction between the pair of solvophobic blocks is attractive. For smaller $\beta\Delta\mu$, i.e.\ for states nearer to coexistence, the solvent mediated potential $W(x_G)$ is longer ranged; the gas intrusion between the blocks lowers the free energy out to larger separations. Close inspection of Fig.~\ref{fig:FMT2DT80Omega_xGapS1E0} shows that there are actually two solution branches to the grand potential. For $\beta\Delta\mu \gtrsim 0.4$ there is only a single smooth branch (not shown). When there are two branches, the one at large $x_G$ corresponds to the case when the density between the blocks is liquid-like and the other, at smaller $x_G$, to when there is a gas-like intrusion. Where the branches meet corresponds to the value of $x_G$ where the evaporation transition occurs for a given $\beta\Delta\mu$. The solvent mediated force between the blocks jumps at the transition. {Note that the potential $W(x_G)$ in Fig.~\ref{fig:FMT2DT80Omega_xGapS1E0} for {\em finite} size blocks (i.e.\ finite $b$) is very different from the corresponding potential between two infinite planar walls (i.e.\ $b\to\infty$). For example, from Fig.~\ref{fig:FMT2DT80Omega_xGapS1E0} we see that when $\beta\Delta\mu=0.05$ the two branches in $W(x_G)$ meet at $x_G\approx8\sigma$. In contrast, for the infinite walls at the same $\beta\Delta\mu$, the two branches meet at $x_G\approx21\sigma$.}

In the same manner used to derive the Kelvin equation \eqref{eq:Kelvin's}, we can use macroscopic thermodynamics to obtain a simple estimate for $W(x_G)$. The grand potential of the system with no blocks present is $\Omega_0 = -p_lV$, where $p_l$ is the pressure of the bulk liquid and $V$ is the volume of the system. The grand potential of the system with one block present in the liquid is
\begin{equation}
	\Omega_1 = -p_l(V-ab^*c^*) + 2(ac^*+ab^*)\gamma_{wl} + 4aE_{l}
	\label{eq:OmegaOneBlock}
\end{equation}
where, $a$, $b^*$, $c^*$ are the effective dimensions of the block, as illustrated in Fig.~\ref{fig:TwoBlock}. Note that $b^*c^*>bc$ is the effective cross sectional area of the block, which includes the fluid exclusion region around the blocks, as discussed below Eq.~\eqref{eq:pot_int_blocks}. Thus $(V-ab^*c^*)$ is the volume occupied by the liquid. Recall that we assume the block is long, i.e.\ $a \to \infty$. $2(ac^*+ab^*)$ is the surface area of the block in contact with the liquid and $\gamma_{wl}$ is the planar wall-liquid interfacial tension. $E_{l}$ is a free energy per unit length so that the final term in Eq.~\eqref{eq:OmegaOneBlock} is the line-tension-like contribution to the grand potential arising from the four edges of the block (corners on the cross-section in Fig.~\ref{fig:TwoBlock}) in contact with the liquid.

Similarly, we can estimate the grand potential when there are two blocks present. If the pair of blocks are close enough together (see e.g.\ the density profile for $\beta\Delta\mu = 0.01$ in Fig.~\ref{fig:FMT2D:PM3D_Rho}) there is a portion of `gas' phase between the blocks, so the grand potential is
\begin{align}
	\Omega_2 = & -p_l(V-2ab^*c^*-ab^*x_G^*) - p_gab^*x_G^*\nonumber\\
	  &+ (4ac^* +2ab^*)\gamma_{wl} + 2ab^*\gamma_{wg} + 2ax_G^*\gamma_{gl}\nonumber\\
			& + 4aE_{l} + 4aE_{gl},
	\label{eq:OmegaTwoBlock} 
\end{align}
where $p_g$ is the pressure of the gas at the same chemical potential as the (bulk) liquid. $x_G^*$ is the effective thickness of the `gas' region between the blocks and as when implementing the Kelvin equation \eqref{eq:Kelvin's}, we set $x_G^*= x_G - 2\sigma$. $\gamma_{wg}$ is the planar wall-gas interfacial tension, $\gamma_{gl}$ is the planar gas-liquid interfacial tension and $E_{gl}$ is the free energy per unit length contribution, i.e.\ the final term in Eq.~\eqref{eq:OmegaTwoBlock} is due to the inner edges of the blocks connecting to a gas-liquid interface. Hence, from Eqs.~\eqref{eq:OmegaOneBlock}, \eqref{eq:OmegaTwoBlock} and \eqref{eq:Young's} the solvent mediated potential, $W(x_G^*) =( \Omega_2 -2 \Omega_1 + \Omega_0) / a$, is given by
\begin{equation}
	W(x_G^*) \approx E + 2b^*\gamma_{lg}\cos\theta + [ 2\gamma_{gl} + b^*(\rho_l - \rho_g)\Delta\mu ]x_G^*,
	\label{eq:SolventMediatedPotentialTwoBlock}
\end{equation}
where $E = 4(E_{gl} - E_{l})$. We have used the standard Taylor expansion of the pressures around the value at gas-liquid bulk coexistence, $p_\text{coex}$, to give $p_l - p_g \approx (\rho_l - \rho_g)\Delta\mu$, where $\rho_l$ and $\rho_g$ are the coexisting bulk liquid and gas densities, respectively. Eq.~\eqref{eq:SolventMediatedPotentialTwoBlock} predicts that the solvent mediated potential is linear in the distance between the blocks $x_G^*$, and thus the force $-\partial W/\partial x_G^* = -2\gamma_{gl} - b^*(\rho_l - \rho_g)\Delta\mu$ is constant when there is a gas-like state between the blocks. The result from Eq.~\eqref{eq:SolventMediatedPotentialTwoBlock}, with $E=0$, is displayed as the thin dotted lines in Fig.~\ref{fig:FMT2DT80Omega_xGapS1E0} for the two extreme cases, $\beta\Delta\mu = 0.05$ and $0.2$. One can see that the gradient of $W(x_G)$ predicted by Eq.~\eqref{eq:SolventMediatedPotentialTwoBlock} agrees very well with the DFT results. However, each line is shifted vertically relative to the DFT curve. This is probably the consequence of having neglected the unknown contribution from the edges, $E$. The difference between the result from  Eq.~\eqref{eq:SolventMediatedPotentialTwoBlock} and the DFT implies that $|E| < 0.5 k_B T/\sigma$. Note that the force $-\partial W/\partial x_G^*$ does not depend on $E$, nor on $\cos\theta$. That the macroscopic thermodynamic result in Eq.~\eqref{eq:SolventMediatedPotentialTwoBlock} agrees rather well with the microscopic DFT results might, once again, come as a surprise to some readers, bearing in mind the microscopic cross-sectional size of the blocks and that the distance between these is only a few solvent particle diameters. The validity of Eq.~\eqref{eq:SolventMediatedPotentialTwoBlock} is partly due to the fact that the correlation length in the intruding gas state is rather short, but this kind of agreement between results of microscopic DFT and simple macroscopic thermodynamic estimates has been observed previously for related problems; see e.g.\ Refs.~\onlinecite{archer2003solvent, archer2005solvent, hopkins2009solvent, malijevsky2015bridging}. Note that the condition $W(x_G^*) = 0$ in Eq.~\eqref{eq:SolventMediatedPotentialTwoBlock} yields
\begin{equation}
x_G^*=\frac{-2\gamma_{lg}\cos\theta}{(\rho_l-\rho_g)\Delta\mu+2\gamma_{lg}/b^*}
\end{equation}
for the separation at which capillary evaporation occurs for identical blocks, i.e.\ the `gas' is thermodynamically stable relative to the liquid for smaller separations. This is a particular case of the formula introduced by Lum and Luzar.\cite{lum1997pathway} In the limit $b^*\to\infty$, the solvent mediated force per unit area is constant, equal to $(\rho_l - \rho_g)\Delta\mu$, in the `gas'. The same result is valid for $\Delta\mu\to0$, in the condensed `liquid' in the case of capillary condensation.\cite{evans1987phase}

\subsection{Two Solvophilic Blocks}

\begin{figure}[t]
	\includegraphics[width=0.49\textwidth]{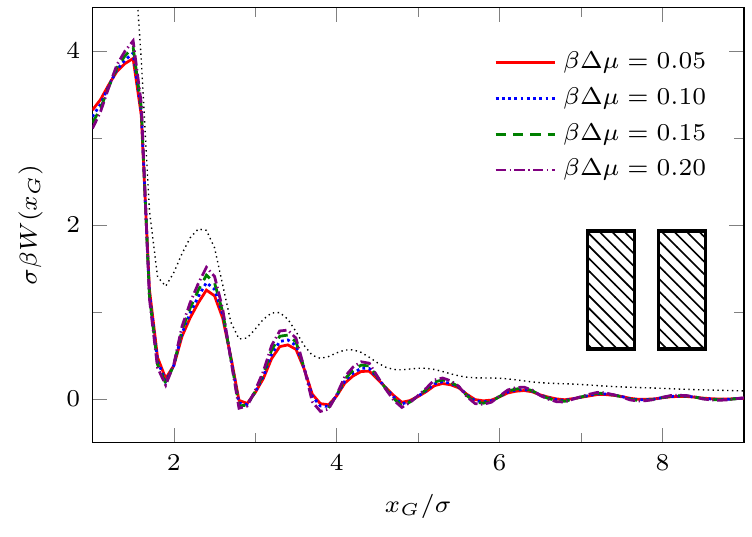}
	\caption{The solvent mediated potential (excess grand potential) between a pair of solvophilic blocks, as a function of distance between blocks $x_G$ for fixed $T=0.8\,T_c$ and $\beta\e_{wf} = 1$. Note that the contact angle for this choice of parameters is $\theta \approx 44^\circ$. {The black dotted line is the corresponding potential per unit area between infinite walls multiplied by $b$, the height of the blocks.}}
	\label{fig:FMT2DT80Omega_xGapS1E1}
\end{figure} 

So far we have discussed the properties of an identical pair of solvophobic blocks. Now we increase the parameter $\e_{wf}$ so that the surface of the blocks attracts more strongly the liquid, i.e.\ the surfaces of the blocks become solvophilic. We set $\beta\e_{wf} = 1$, which for the planar wall has the contact angle $\theta = 43.7^\circ$, see Fig.~\ref{fig:FMT1DT80ThetaEw}. The density profiles for the blocks of the same dimensions (not displayed) are, for all values of $\beta\Delta\mu$, qualitatively similar to the profile corresponding to $\beta\Delta\mu = 0.4$ in Fig.~\ref{fig:FMT2D:PM3D_Rho}, but with higher densities at the surface of the blocks and larger amplitude oscillations in the density profile around the blocks. The same is true for the compressibility. The key difference between a pair of solvophobic blocks and a pair of solvophilic blocks is that there is no capillary evaporation of the liquid in the gap between the solvophilic blocks as $\beta\Delta\mu\to0$. This has profound consequences for the solvent mediated potential.

Fig.~\ref{fig:FMT2DT80Omega_xGapS1E1} shows the solvent mediated potential $W(x_G)$ between the solvophilic blocks. We see pronounced oscillations as the distance between the blocks is decreased. Also, since $W(x_G)$ decreases (albeit in damped oscillatory fashion) as $x_G$ is increased, this indicates that the effective interaction potential between a pair of solvophilic blocks is repulsive. Note that $W(x_G)$ is almost independent of $\beta\Delta\mu$ in this particular case. The results in Fig.~\ref{fig:FMT2DT80Omega_xGapS1E1} are quite similar to those obtained for two planar walls with the same $\beta\e_{wf}$ ({thin dotted black line}). Note that for planar walls the asymptotic decay, $L\to\infty$, of the excess grand potential per unit area $W(L)$ is known~\cite{attard1991interaction, maciolek2004solvation} for various choices of the fluid-fluid and wall-fluid potentials. For our present choice [Eqs.~\eqref{eq:attLJ} and \eqref{eq:FMT1D:Ew}], with $\beta\e_{wf}=1$, theory predicts $\beta W(L) \sim 0.934 L^{-2}$, as $L \to \infty$, i.e.\ the solvent mediated force per unit area $-(\partial W / \partial L)_{T,\mu}$ is repulsive and decays $\sim L^{-3}$. We are not able to investigate the asymptotics numerically for blocks.

\subsection{One solvophobic and one solvophilic block and patchy blocks}

\begin{figure*}[t]
	\includegraphics[width=0.4\textwidth]{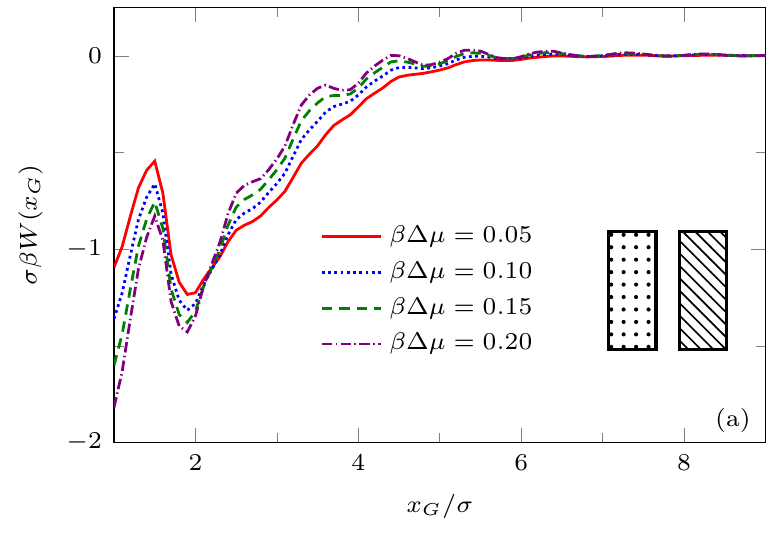}
	\includegraphics[width=0.4\textwidth]{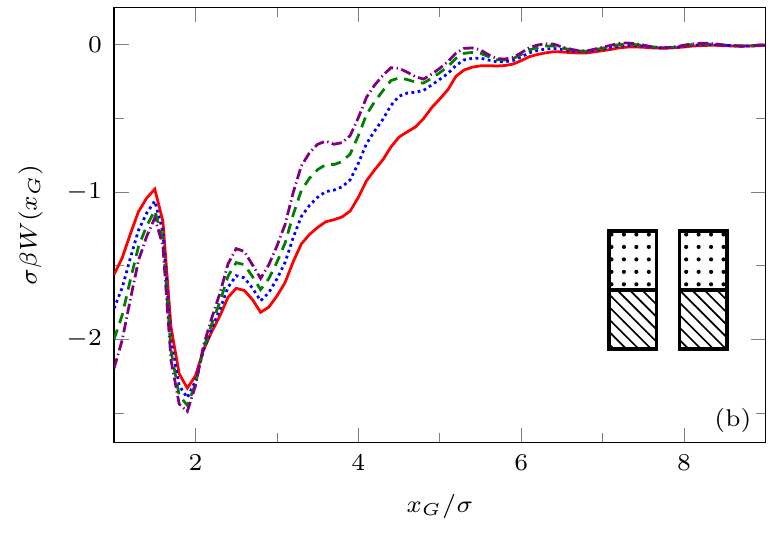}
	\includegraphics[width=0.4\textwidth]{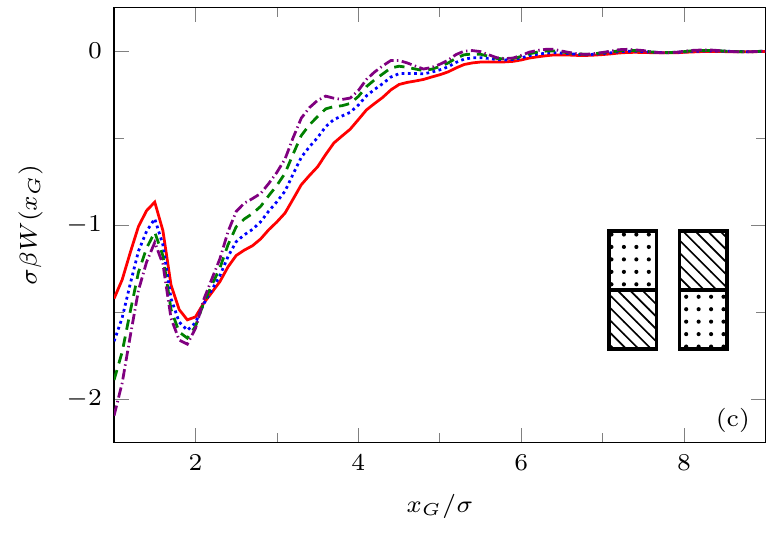}
	\includegraphics[width=0.4\textwidth]{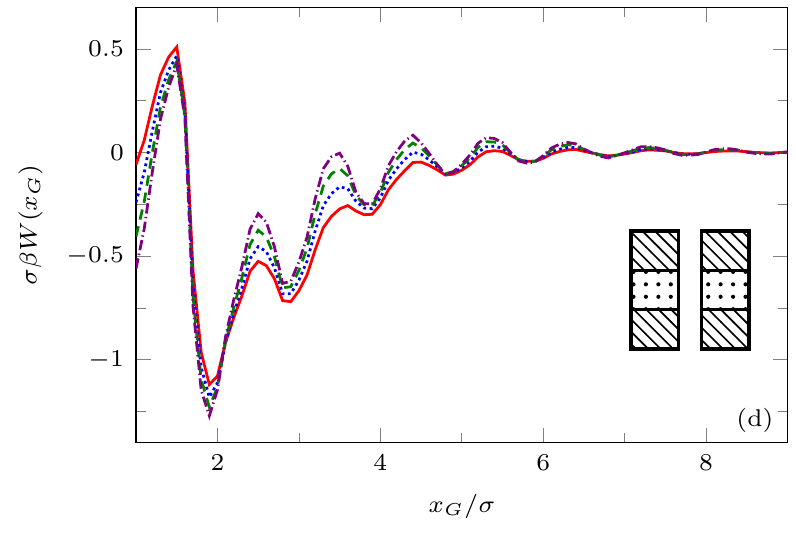}
	\includegraphics[width=0.4\textwidth]{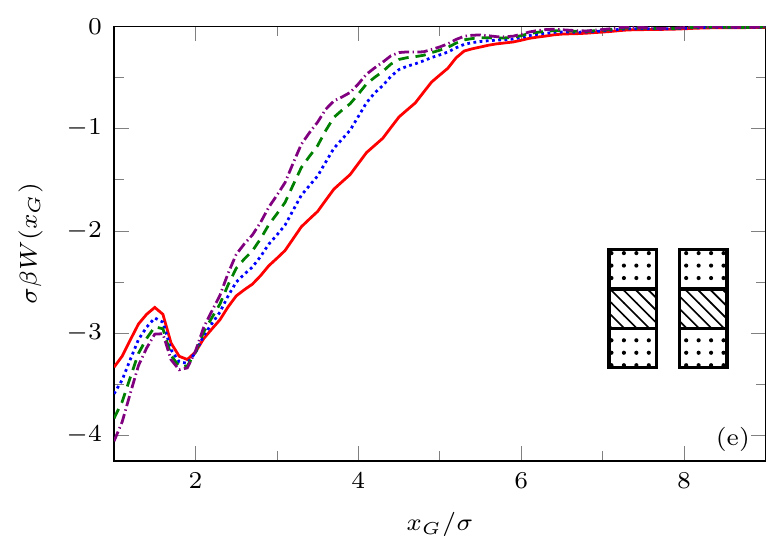}
	\includegraphics[width=0.4\textwidth]{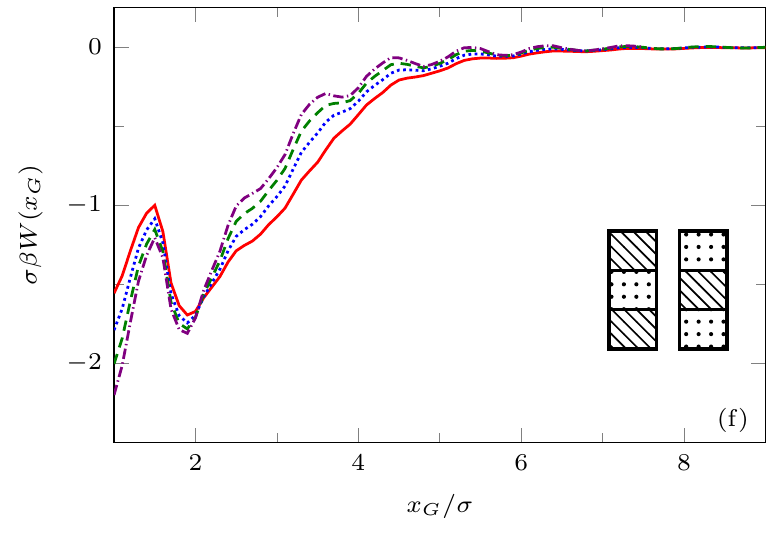}

	\caption{The solvent mediated potential (excess grand potential) between various pairs of blocks as a function of distance between blocks, $x_G$, for fixed $T=0.8\,T_c$. The structure of each block is specified by the inset where dots represent solvophobic areas ($\beta\e_{wf} = 0.3$) and the diagonal lines represent solvophilic areas ($\beta\e_{wf} = 1$). When patchy blocks are aligned the same way, we call this `even' alignment, otherwise we refer to this as `odd'. Thus, in (b) there is even and in (c) odd alignment. For (d)-(f) the block is split into three segments; (d) and (e) are even while (f) is odd.}
	\label{fig:FMT2DT80Omega_xGapRest}
\end{figure*} 

\begin{figure*}[t]
	\includegraphics[width=0.99\textwidth]{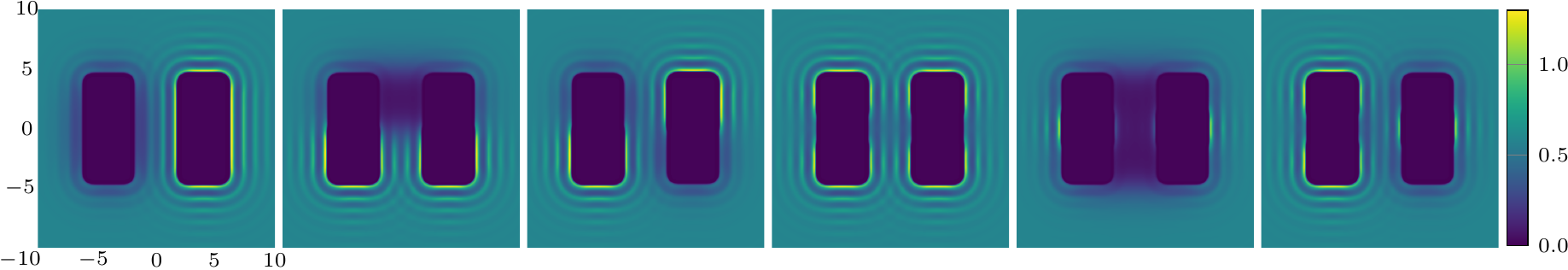}
	\\[0.5em]
	\includegraphics[width=0.99\textwidth]{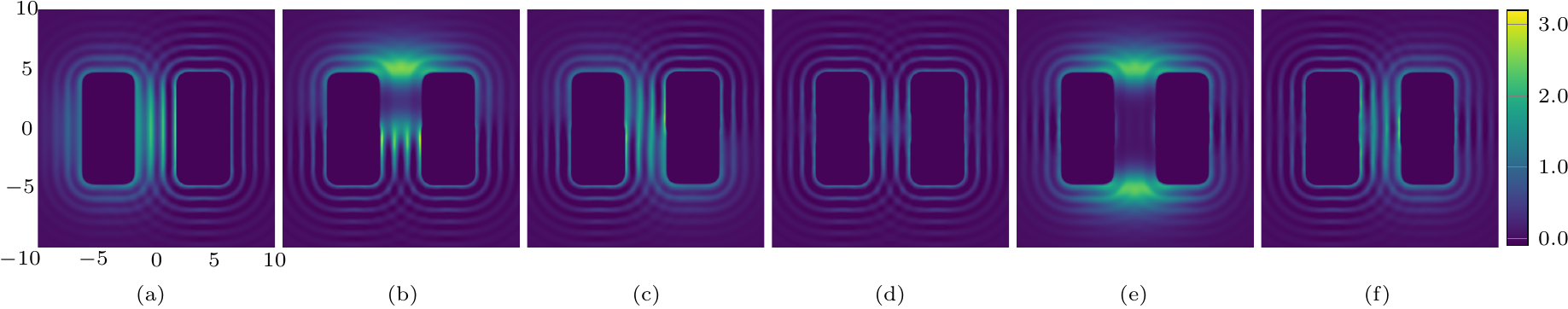}
	\caption{Liquid density profiles $\sigma^3\rho(x,y)$ (top) and the local compressibility $k_BT\sigma^3\chi(x,y)$ (bottom) around different pairs of blocks for the temperature $T =0.8\,T_c$ and chemical potential $\beta\Delta\mu = 0.01$. In all cases the blocks are a distance $x_G = 5\sigma$ apart. The labels (a)-(f) refer to the same pair of blocks as described in Fig.~\ref{fig:FMT2DT80Omega_xGapRest}.}
	\label{fig:FMT2D:PM3D_RhoRest}
\end{figure*}

The two previous subsections discuss the solvent mediated interactions $W(x_G)$ between pairs of blocks that are identical. We now present results for $W(x_G)$ for the case when one of the blocks is solvophobic and the other is solvophilic. We also consider various pairs of block having a mixture of solvophobic and solvophilic patches. We split each block into a maximum of three segments. The DFT results for the solvent mediated potentials between the various blocks are shown in Fig.~\ref{fig:FMT2DT80Omega_xGapRest}, with the inset giving a sketch of the arrangement of the patches: dotted regions are solvophobic and diagonally striped regions are solvophilic. In all cases in Fig.~\ref{fig:FMT2DT80Omega_xGapRest}, we notice that there is a local minimum of $W(x_G)$ occurring when $x_G\approx2\sigma$. This is the distance at which the two exclusion zones around the blocks meet, so that for $x_G$ less than this value, the fluid density between the blocks is almost zero. In general, the range of the solvent mediated interaction decreases as $\beta\Delta\mu$ is increased. Note that having blocks with only one solvophobic segment causes the solvent mediated potential $W(x_G)$ to become attractive. Nevertheless, $W(x_G)$ retains the oscillatory behaviour of the pure solvophilic blocks observed in Fig.~\ref{fig:FMT2DT80Omega_xGapS1E1}. Furthermore, the oscillations in the potential are enhanced when the solvophilic patches are together on the ends of the blocks -- see Fig.~\ref{fig:FMT2DT80Omega_xGapRest}(d). In Fig.~\ref{fig:FMT2D:PM3D_RhoRest} we display a series of density profiles and local compressibility profiles corresponding to all the cases displayed in Fig.~\ref{fig:FMT2DT80Omega_xGapRest}. We observe that whenever two solvophobic segments are opposite one another, a gas-like region forms between the blocks provided these are sufficiently close (as they are in Fig.~\ref{fig:FMT2D:PM3D_RhoRest}) and this leads to large values of the local compressibility $\chi(\vec{r})$ in these regions.

It is particularly instructive to compare the results in Figs.\ \ref{fig:FMT2DT80Omega_xGapRest}(e) and \ref{fig:FMT2D:PM3D_RhoRest}(e), corresponding to two solvophobic patches facing each other at both ends of the blocks, with the corresponding ones for identical uniform solvophobic blocks, Figs.\ \ref{fig:FMT2D:PM3D_Rho}, \ref{fig:FMT2D:PM3D_Chi} and \ref{fig:FMT2DT80Omega_xGapS1E0}. For $\beta\Delta\mu=0.05$, the solvent mediated potential in Fig.\ \ref{fig:FMT2DT80Omega_xGapRest}(e) has a form close to that in Fig.\ \ref{fig:FMT2DT80Omega_xGapS1E0}. The separation, $x_G\approx 5\sigma$, at which capillary evaporation occurs is smaller for the patchy case than for the uniform case, $x_G\approx 8\sigma$. However, in both cases one finds a linear solvent mediated potential at smaller separations with constant gradients; the magnitude of the force is similar in both cases. Such behaviour is consistent with reduced area of (facing) solvophobic regions. Recall that for two identical blocks Eq.~\eqref{eq:SolventMediatedPotentialTwoBlock} implies that the force does not depend on $\cos\theta$.

\subsection{Blocks shifted vertically}

\begin{figure*}[t]
	\includegraphics[width=0.495\textwidth]{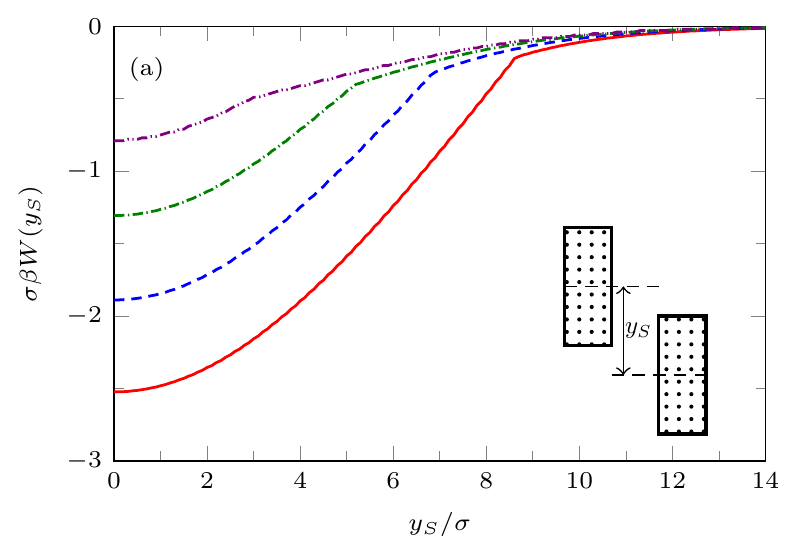}
	\includegraphics[width=0.495\textwidth]{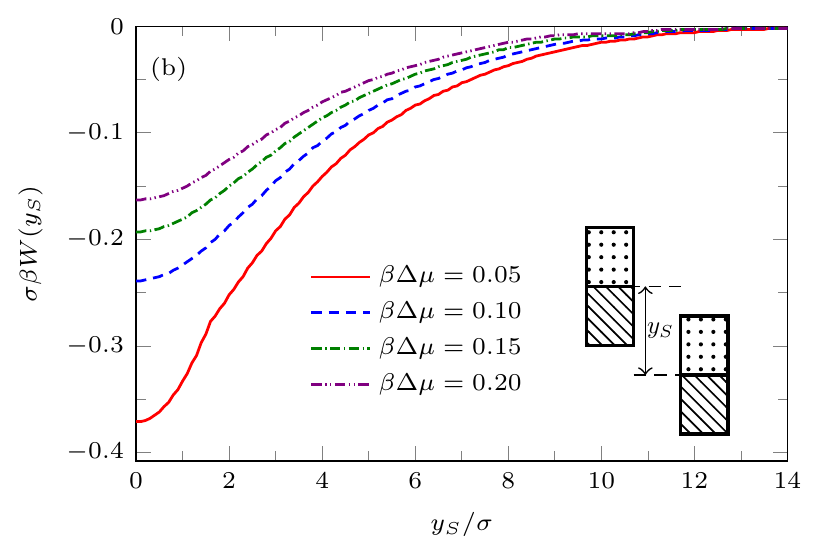}
	\caption{The solvent mediated potential (excess grand potential) between (a) an identical pair of solvophobic blocks and (b) a pair of patchy blocks divided into two segments: half solvophobic and half solvophilic, aligned evenly, as a function of $y_S$, the vertical distance between the horizontal lines through the block mid-points -- see inset. For the solvophobic segment $\beta\e_{wf} = 0.3$ and for the solvophilic segment $\beta\e_{wf}=1$. $T=0.8\,T_c$ and $x_G=5\sigma$. Note that in (a) and (b) there are two branches -- see text.}
	\label{fig:FMT2DT80Omega_yShift}
\end{figure*}

The results presented in the previous subsections are for the case when the centres of the blocks are at $y=0$ and only the distance between the closest faces $x_G$ is varied. Now we fix $x_G=5\sigma$ and move one of the blocks vertically along the $y$-axis [cf.\ inset of Fig.~\ref{fig:FMT2DT80Omega_yShift}]. The vertical distance from the $x$-axis is defined as $y_S$ (in the previous subsections $y_S=0$). The solvent mediated potential $W(y_S)$ for a pair of solvophobic blocks and a pair of patchy blocks (divided into two segments: half solvophobic and half solvophilic, aligned evenly) is shown in Fig.~\ref{fig:FMT2DT80Omega_yShift}. In both cases we see that $W(y_S)$ is attractive, with a minimum at $y_S=0$. This indicates that the preferred position (lower grand potential) is when the pair of blocks are aligned, with $y_S=0$. We also see that for a given chemical potential the range and depth of the potential is greater for a pair of fully solvophobic blocks [Fig.~\ref{fig:FMT2DT80Omega_yShift}(a)] than for a pair of two-segment blocks aligned evenly [Fig.~\ref{fig:FMT2DT80Omega_yShift}(b)]. This is as one would expect, since the amount of solvophobic area on each block is greater in the former case (a). For a pair of solvophobic blocks, we showed in Fig.~\ref{fig:FMT2DT80Omega_xGapS1E0} that the solvent mediated potential $W(x_G)$ varies approximately linearly with $x_G$, on the `gas branch' arising for smaller values of $x_G$. However, we see from Fig.~\ref{fig:FMT2DT80Omega_yShift}(a) that for fixed $x_G$ the solvent mediated potential is not a linear function of $y_S$. For the pair of two-segment blocks aligned evenly (Fig.~\ref{fig:FMT2DT80Omega_yShift}(b)) we do not see any oscillations in the solvent mediated potential as $y_S$ is varied -- recall that there are oscillations as $x_G$ is varied -- see Fig.~\ref{fig:FMT2DT80Omega_xGapRest}(b). Close inspection of Fig.~\ref{fig:FMT2DT80Omega_yShift} shows that within the present mean-field DFT approach there are actually two solution branches to the grand potential for both types of blocks. The branch for large $y_S$ corresponds to a liquid-like density between the blocks while the other branch at smaller $y_S$, corresponds to the density between the blocks being gas-like. Consistent with our earlier discussion, the evaporation transition occurs at the value of $y_S$ where the two branches meet and the solvent mediated force between the blocks jumps at this point. The value of $y_S$ at which this transition occurs varies with $\beta\Delta\mu$.

{Note that it is straightforward to derive a formula for $W(y_s)$ analogous to that in Eq.~\eqref{eq:SolventMediatedPotentialTwoBlock}, making the same assumptions. However, the assumption that the gas-liquid interface meets the blocks at the corners is no longer necessarily true and the resulting formula gives poor agreement with the DFT.}

\subsection{Blocks at an angle}

\begin{figure*}[t]
	\includegraphics[width=0.99\textwidth]{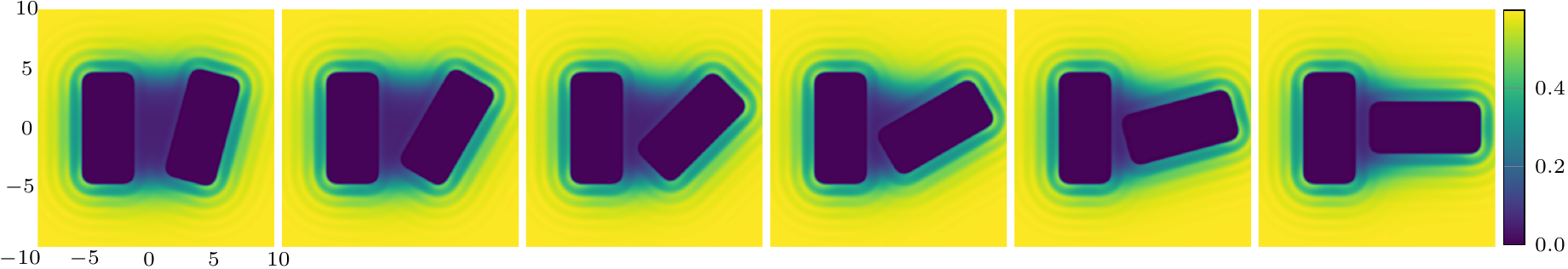}
	\\[0.5em]
	\includegraphics[width=0.99\textwidth]{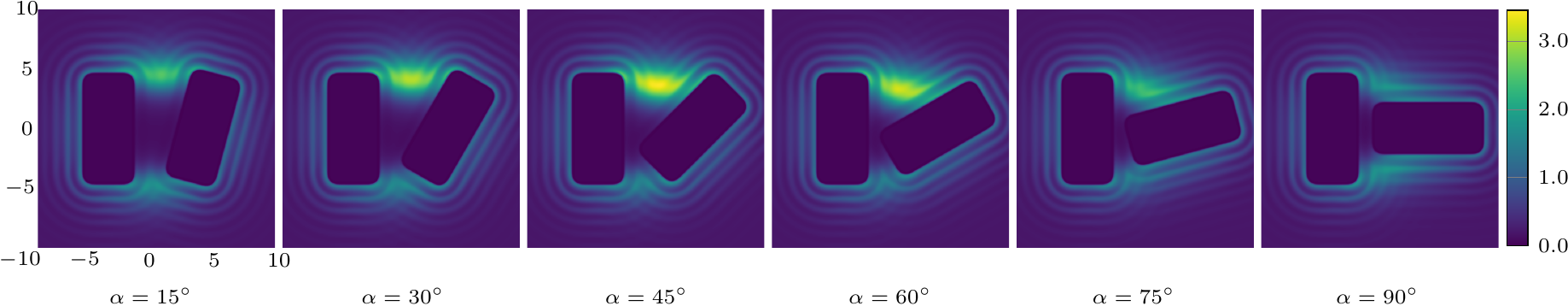}
	\caption{Liquid density profiles $\sigma^3\rho(x,y)$ (top) and the local compressibility $k_BT\sigma^3\chi(x,y)$ (bottom) around a pair of (identical) solvophobic blocks, for fixed temperature $T=0.8\,T_c$ and chemical potential $\beta\Delta\mu = 0.05$. The distance between the centre of blocks is fixed: $x_C = 8\sigma$ and the relative orientation angle $\alpha$ is varied. The density and compressibility profiles for $\alpha=0$ are not shown since these form part of the sequence in Figs.~\ref{fig:FMT2D:PM3D_Rho} and \ref{fig:FMT2D:PM3D_Chi}.}
	\label{fig:FMT2DPM3D_xC8A}
\end{figure*}

So far we have considered pairs of blocks with their faces aligned parallel to each other. We now consider a pair of identical solvophobic blocks with second block rotated by an angle $\alpha$ with respect to the centre of the first, i.e.\ $\alpha$ is the angle between the orientation vectors of the two blocks. In Fig.~\ref{fig:FMT2DPM3D_xC8A} we plot the density and compressibility profiles as the angle $\alpha$ is varied whilst keeping the distance between the centres of the blocks fixed, $x_C = 8\sigma$ (note that $x_C \ne x_G$). The temperature $T=0.8\,T_c$ and chemical potential $\beta\Delta\mu = 0.05$ are also fixed. We present results for a range of angles; by symmetry we only need to consider the range $0^\circ\leq\alpha\leq90^\circ$.

Fig.~\ref{fig:FMT2DPM3D_xC8A} (top) shows that as $\alpha$ is increased for fixed $x_C=8\sigma$, the gas-like region between the blocks remains. The area of one of the interfaces between the gas-like region and the bulk liquid increases, while the other decreases. Additionally, we see that the volume of the gas-filled region between the blocks decreases as $\alpha$ is increased from zero, since the blocks become closer to each other. {Note also that for the larger values of $\alpha$, the gas-liquid interface does not connect to the corners of the blocks, which must be taken into account if generalising Eq.~\eqref{eq:SolventMediatedPotentialTwoBlock} to derive an approximation for $W$ as a function of $\alpha$.} From the corresponding compressibility profiles in Fig.~\ref{fig:FMT2DPM3D_xC8A} (bottom) we see that $\chi(\vec{r})$ is largest in the gas-liquid interfaces, as previously. Also, the peak value of the compressibility increases as $\alpha$ is increased from zero, attaining its maximum value when $\alpha\approx45^\circ$. Increasing $\alpha$ further leads to a drop in the peak value of the compressibility.

\begin{figure}[t]
	\includegraphics[width=0.495\textwidth]{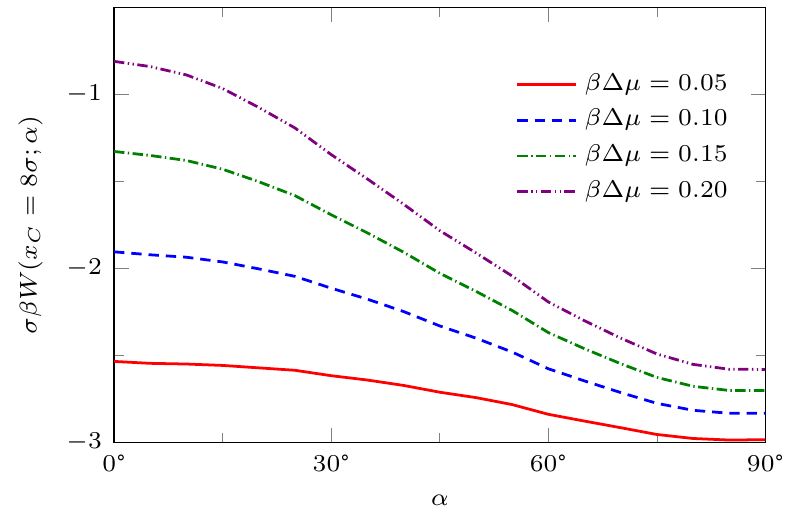}
	\caption{The solvent mediated potential (excess grand potential) between a pair of solvophobic blocks for various values of chemical potential as a function of the relative orientation angle, $\alpha$. As in Fig.~\ref{fig:FMT2DPM3D_xC8A} we fix the centres of the pair of blocks such that the distance between the centres $x_C = 8\sigma$ and rotate one of the blocks by $\alpha$. $T=0.8\,T_c$ and $\beta\e_{wf} = 0.3$. }
	\label{fig:FMT2DT80Omega_AnglexC8S1E0}
\end{figure}

In Fig.~\ref{fig:FMT2DT80Omega_AnglexC8S1E0} we plot the solvent mediated potential for two solvophobic blocks as a function of $\alpha$ for fixed distance between the centres of the blocks, $x_C = 8\sigma$, corresponding to the profiles in Fig.~\ref{fig:FMT2DPM3D_xC8A}. We see that the minimum of the solvent mediated potential occurs when $\alpha=90^\circ$ for fixed $x_C = 8\sigma$. This is because as the angle is varied, the blocks become closer to each other as $\alpha \to 90^\circ$ (see Fig.~\ref{fig:FMT2DPM3D_xC8A}) and this leads to the excess grand free energy being lower. However, if we rotate the solvophobic blocks {\em and} also move the centres of the blocks such that closest distance between the two blocks $x_G$ is always constant, we find the minimum of the grand potential is when $\alpha = 0^\circ$ (not shown). In this case, it is because rotating to $\alpha = 90^\circ$ results in a smaller area of the block surfaces being opposite one another than when $\alpha = 0^\circ$. Generically the attractive well in the solvent mediated potential between the blocks becomes deeper (i.e.\ stronger attraction) as $\beta\Delta\mu \to 0$.

\begin{figure}[t]
	\includegraphics[width=0.495\textwidth]{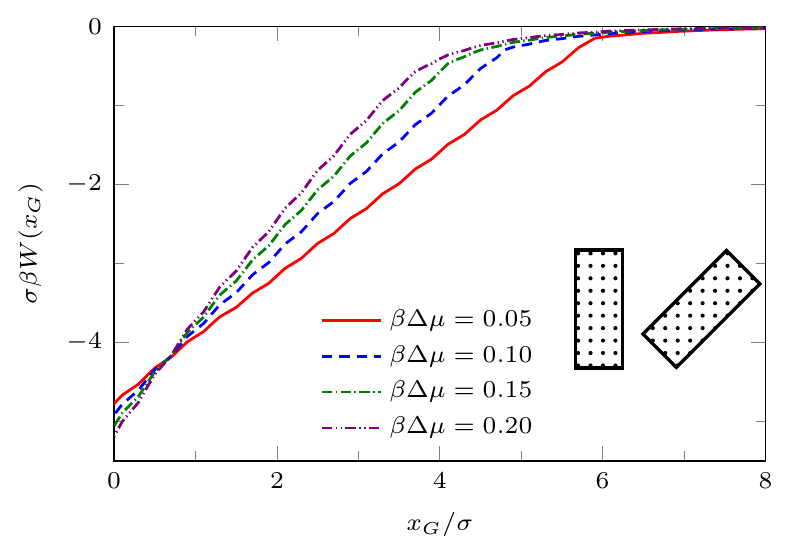}
	\caption{The solvent mediated potential (excess grand potential) between a pair of solvophobic blocks for various values of the chemical potential as a function of the distance between the closest points of the blocks, $x_G$, with one of the blocks rotated at fixed angle $\alpha = 45^\circ$. $T=0.8\,T_c$ and $\beta\e_{wf} = 0.3$. Note that for small values of $\beta\Delta\mu$ there are two branches -- see text.}
	\label{fig:FMT2DT80Omega_xCentreS1E0_45}
\end{figure}

In order to analyse further the solvent mediated potential between the solvophobic blocks, we fix the relative orientation between the blocks at $\alpha=45^\circ$ and vary the separation between the blocks $x_G$, which is the distance between the closest points on the pair of blocks. $W(x_G)$ is shown in Fig.~\ref{fig:FMT2DT80Omega_xCentreS1E0_45} for fixed temperature $T=0.8\,T_c$ and wall attraction $\beta\e_{wf} = 0.3$. In the inset we sketch the relative orientations of the two blocks. Thus, $x_G$ is the distance from the left-most corner of the right hand block to the near face of the left hand block. From Fig.~\ref{fig:FMT2DT80Omega_xCentreS1E0_45}, we see that the solvent mediated potential between the pair of solvophobic blocks with fixed $\alpha=45^\circ$ is qualitatively similar to that for $\alpha=0^\circ$, see Fig.~\ref{fig:FMT2DT80Omega_xGapS1E0}. For small $\beta\Delta\mu$, i.e.\ for states nearer to coexistence, the solvent mediated potential $W(x_G)$ is longer ranged (although not as long-ranged as when the faces are parallel, $\alpha=0^\circ$, shown in Fig.~\ref{fig:FMT2DT80Omega_xGapS1E0}) and also has two solution branches to the grand potential. The branch for large $x_G$ corresponds to the liquid-like density between the blocks and the other, at smaller $x_G$, is when the density between the blocks is gas-like. Once again the evaporation transition occurs at the value of $x_G$ where the two branches cross and the solvent mediated force jumps at this value of $x_G$ for the given $\beta\Delta\mu$. 

\section{Concluding remarks}
\label{sec:Conclusions}

Using classical DFT we have calculated the liquid density profile and the local compressibility around pairs of solvophobic, solvophilic and patchy blocks immersed in a simple LJ like solvent. We have also calculated an important thermodynamic quantity, namely the solvent mediated interaction potential between the blocks $W(x_G)$. When both blocks are solvophobic, the potential $W(x_G)$ is an almost linear function at small separations $x_G$, is strongly attractive and is very sensitive to the value of $\beta\Delta\mu$; see Fig.~\ref{fig:FMT2DT80Omega_xGapS1E0}. In this regime, treating the system using macroscopic thermodynamics, i.e.\ using  Eq.~\eqref{eq:SolventMediatedPotentialTwoBlock}, turns out to be a rather good approximation for $W(x_G)$. Although this may seem surprising, given that the blocks we consider have the microscopic cross sectional area $\approx 10\sigma \times 5\sigma$, it is in keeping with recent simulation studies\cite{kanduc2016water, jabes2016universal} of water induced interactions between hydrophobes. In contrast, when both blocks are solvophilic, the potential $W(x_G)$ is oscillatory but overall repulsive and exhibits only a weak dependence on $\beta\Delta\mu$; see Fig.~\ref{fig:FMT2DT80Omega_xGapS1E1}. When the blocks are patchy, the nature of the solvent mediated potential is complex. However, we find that if solvophobic patches are present, are sufficiently large and near to one another (facing each other on the opposing blocks), then their contribution to the effective potential dominates (see Fig.\ \ref{fig:FMT2DT80Omega_xGapRest}). Then the potential $W(x_G)$ is still strongly attractive and is nearly linear in $x_G$ for small $\beta\Delta\mu$, particularly if the solvophobic patches are on the ends of the blocks [see Fig.\ \ref{fig:FMT2DT80Omega_xGapRest}(e)]. From Fig.\ \ref{fig:FMT2DT80Omega_yShift} we see that for fixed $x_G$ there is a minimum in $W$ as a function of the vertical distance $y_s$, when the solvophobic patches on the blocks are aligned. 

For a pair of identical solvophobic blocks, the solvent mediated potential per unit length of the blocks is $\approx-5k_BT$ when the blocks are close to contact (see Fig.\ \ref{fig:FMT2DT80Omega_xGapS1E0}). Thus, if we assume that the blocks are actually finite in length, with length $a=10\sigma$ (i.e.\ finite blocks of size $10\sigma \times 10\sigma \times 5\sigma$), then when the blocks are close to contact we have $W(x_G \lesssim \sigma)\approx -50k_BT$ or about $-120$ kJ mol$^{-1}$ at ambient temperature. This is the same order of magnitude as the solvent mediated potentials between a hydrophobic (polymeric) solute of a similar size and a hydrophobic SAM surface measured in computer simulations employing a realistic model of water -- see Fig.~6(c) in Ref.\ \onlinecite{jamadagni2011hydrophobicity} and also Ref.\ \onlinecite{jamadagni2009surface}. Moreover, it is important to note that when the SAM surface is strongly hydrophobic, the solvent mediated potentials in Ref.\ \onlinecite{jamadagni2011hydrophobicity} display a portion that is almost linear. Hydrophobic interactions also play a role in determining the structure of proteins: simulations suggest capillary evaporation between hydrophobic patches can lead to strong forces between protein surfaces.\cite{liu2005observation} Given these observations, we expect that the results described here for a simple LJ like liquid incorporate the essential physics of a realistic model of a water solvent.

In the vicinity of a single solvophobic surface the solvent density is lower, when $\beta\Delta\mu$ is sufficiently small. However, the thickness of the depleted layer is only one or two particle diameters -- see Fig.\ \ref{fig:FMT1DT80Ew03} corresponding to $\theta \approx 144^\circ$. This is consistent with the x-ray studies of water at a water-OTS (octadecyl-trichlorosilane) surface reported in Ref.\ \onlinecite{mezger2006high} and with simulation results for SPC/E water at non-polar substrates.\cite{janecek2007interfacial} When two solvophobic surfaces become sufficiently close a gas-like region forms between the blocks. The extent of this can be large, see e.g.\ Fig.~\ref{fig:FMT2D:PM3D_Rho}, and the density profile passing from the gas inside to the liquid outside of the blocks closely resembles the free gas-liquid interfacial profile -- see Fig.\ \ref{fig:FMT2D:E3:Ew05:xGap7:P2}. Moreover, the local compressibility is large in the neighbourhood of this interface, indicating that it is a region with large density fluctuations. Given that this interface is pinned to the corners of the blocks -- see Fig\ \ref{fig:FMT2D:PM3D_Rho} -- we do not expect significant ``capillary wave'' broadening of the profile beyond the present mean-field DFT, as one would normally expect at a macroscopic free interface.

As the separation between solvophobic blocks is increased, there is a jump in the solvent mediated force when the blocks reach a particular distance, $x_G=x_J$, where the state minimising the grand potential changes from one with a gas-like density between the blocks to one where this is liquid-like. Within DFT the potential $W(x_G)$ has two branches and there is a discontinuity in the the gradient at $x_J$ -- see e.g.\ Fig.~\ref{fig:FMT2DT80Omega_xGapS1E0}. We do not display the metastable portions of the branches of $W(x_G)$; these do not extend very far from the crossing point indicating that the height of the nucleation barrier is small. This is due to the small size of the blocks and the small values of $x_G$. For hydrophobic surfaces with greater surface area and at a greater distance apart, the free energy barrier should be larger; for a recent discussion of nucleation pathways to capillary evaporation in water see Ref.\ \onlinecite{remsing2015pathways}.

We have also studied the local compressibility $\chi(\vec{r})$ in the liquid between and surrounding pairs of blocks of differing nature. The local compressibility exhibits pronounced peaks; these indicate where the local density fluctuations are large. These fluctuations are maximal close to the incipient gas-liquid interface -- see for example the central plot in Fig.\ \ref{fig:FMT2D:PM3D_Chi}, which is for $\beta\Delta\mu=0.22$, and also Fig.\ \ref{fig:FMT2D:PM3D_RhoRest} (b) and (e) for $\beta\Delta\mu = 0.01$. Fig.~\ref{fig:FMT2DPM3D_xC8A} displays how for angled blocks $\chi(\vec{r})$ depends on alignment and the confining geometry. When pronounced fluctuations, in conjunction with a depleted surface density, are observed in simulations of water at hydrophobic interfaces, this phenomenon is often ascribed to the disruption of the water hydrogen bonding network. Given that we observe similar behaviour for a simple LJ like liquid close to solvophobic substrates, we argue that this phenomenon is by no means specific to water. Rather it is due (i) to the weak bonding between the fluid and the (solvophobic) surface and (ii) the system being close to bulk gas-liquid phase coexistence, i.e.\ a small value of $\beta\Delta\mu$. Thus, since the LJ like fluid considered here is representative of a broad class of simple liquids, we expect strong attraction between solvophobic surfaces, enhanced density fluctuations near such surfaces and other features of hydrophobicity to manifest themselves whenever the solvent, whatever its type, is near to bulk gas-liquid phase coexistence. There are obvious advantages, both in simulation and theory, in performing detailed investigations for simple model liquids, especially when tackling subtle questions of surface phase transitions such as critical drying.\cite{evans2016critical}

\section*{Acknowledgements}

We benefitted from useful discussions about this work with Chris Chalmers and Nigel Wilding. BC acknowledges the support of EPSRC and the work of RE was supported by a Leverhulme Emeritus Fellowship: EM-2016-031.
	

%

\end{document}